\documentclass[pra,reprint,amsmath,amssymb,aps,a4paper,longbibliography]{revtex4-2}
\usepackage{graphicx}
\usepackage[utf8]{inputenc}
\usepackage{braket}
\usepackage{comment}
\usepackage{float}
\usepackage{siunitx}
\usepackage[colorlinks=true, linkcolor=blue, urlcolor=blue, citecolor=blue]{hyperref}
\usepackage{lipsum}
\usepackage{cleveref}
\usepackage{bibunits}

\newcommand{\ua}{\uparrow}
\newcommand{\da}{\downarrow}

\begin{document}
\begin{bibunit}[apsrev4-2]
\title{Effects of spin--orbit coupling and in-plane Zeeman fields on the critical current\\ in two-dimensional hole gas SNS junctions}

\author{Jonas Lidal}
\author{Jeroen Danon}
\affiliation{Center for Quantum Spintronics, Department of Physics, Norwegian University of Science and Technology, NO-7491 Trondheim, Norway}

\date{\today}

\begin{abstract}
Superconductor--semiconductor hybrid devices are currently attracting much attention, fueled by the fact that strong spin--orbit interaction in combination with induced superconductivity can lead to exotic physics with potential applications in fault-tolerant quantum computation.
The detailed nature of the spin dynamics in such systems is, however, often strongly dependent on device details and hard to access in experiment.
In this paper we theoretically investigate a superconductor--normal--superconductor junction based on a two-dimensional hole gas with additional Rashba spin orbit--coupling, and we focus on the dependence of the critical current on the direction and magnitude of an applied in-plane magnetic field.
We present a simple model, which allows us to systematically investigate different parameter regimes and obtain both numerical results and analytical expressions for all limiting cases.
Our results could serve as a tool for extracting more information about the detailed spin physics in a two-dimensional hole gas based on a measured pattern of critical currents.
\end{abstract}

\maketitle

\section{\label{sec:intro}Introduction}

Hybrid devices made of superconductors and semiconductors have gained much interest in recent years due to their rich and complex behavior.
Spin--orbit coupling in combination with superconducting correlations induced via the proximity effect can give rise to exotic spin physics inside the semiconductor, which could be exploited to engineer topological superconductivity~\cite{Sau2010,Alicea2010,Oreg2010,Lutchyn2010,Sarma2015,Hell2017,Lutchyn2018a}.
Since such topological superconductors are expected to host low-energy Majorana modes that obey non-Abelian anyonic statistics, they could provide a platform for implementing fault-tolerant quantum computation with topologically protected qubit operations~\cite{Nayak2008,Pachos2012,Stanescu2020}.

Arguably the simplest hybrid device one can create using superconducting and normal elements is the superconductor--normal--superconductor (SNS) junction, which finds applications in a wide range of directions, including superconducting qubits~\cite{Nakamura1999,Friedman2000,Martinis2002,Koch2007,Casparis2018} and electronic and magnetic measuring devices~\cite{Jaklevic1964,Silver1967,Tesche1977,Kautz1996,Kleiner2004}.
In addition to being an essential component of superconducting circuits, an SNS junction can also be used for studying the underlying properties of the constituent elements of the hybrid structure.
For the case of a semiconducting normal region, an SNS setup allows to probe details of the spin--orbit interaction in the semiconductor and its interplay with the Zeeman effect~\cite{Bezuglyi2002,yokoyamaAnomalousJosephsonEffect2014,rasmussenEffectsSpinorbitCoupling2016}, as well as to study phase transitions into and out of topological phases~\cite{San-Jose2013,San-Jose2014,Dominguez2022}.

One quantity that encodes several details of the underlying physics of the system is the critical current through the SNS junction, i.e., the maximal supercurrent the junction can support.
By applying a magnetic field perpendicular to a two-dimensional junction, information about the current density distribution can be extracted from the measured critical current~\cite{Dynes1971}.
For a uniform current distribution, the critical current as a function of the out-of-plane magnetic field emerges as a so-called Fraunhofer pattern, which reflects the flux enclosed by the junction.
A deviation from a Fraunhofer pattern is a sign of a non-uniform current distribution and the pattern of critical current can be directly related to the actual current distribution profile in the junction~\cite{Zappe1975,Hui2014,Hart2014,Pribiag2015,Allen2016,Suominen2017,Kononov2020}.

The field-dependent behavior of an SNS junction is heavily influenced by the properties of the normal part, and junctions based on a wide range of materials have been explored in the past~\cite{Kontos2002,Andersen2006,Kurter2015,Calado2015}.
In this paper we focus on SNS junctions comprised of a two-dimensional hole gas (2DHG) contacted by two conventional superconductors.
Our choice is motivated by the recent surge in interest for lower-dimensional quantum devices hosted in 2DHGs~\cite{Watzinger2018,Ridderbos2019,Hendrickx2020,Froning2021,Scappucci2021,Jirovec2021,Piot2022,Borsoi2022}, which was sparked by their interesting properties including strong inherent and tunable spin--orbit interaction~\cite{Marcellina2017,Philippopoulos2020,Terrazos2021,Bosco2021,Bosco2021a} and highly anisotropic and tunable $g$-tensors~\cite{Crippa2018,Gradl2018,Liles2021,Qvist2022}, all caused by the underlying $p$-type orbital structure of the valence band states~\cite{winkler2003}.
Additionally, germanium-based hole gases have recently shown great promise for straightforward integration with superconducting elements~\cite{Hendrickx2018,Hendrickx2019,Aggarwal2021,Tosato2022}.

The effective spin--orbit interaction and Zeeman coupling that together can give rise to its useful properties depend strongly on many details of the 2DHG, including its exact out-of-plane confining potential, the carrier density, strain, and the local electrostatic landscape.
For this reason it is not always straightforward to access the relevant underlying spin--orbit and $g$-tensor parameters in experiment for a given system.
Here, we theoretically study the dependence of the critical current through a 2DHG-based SNS junction on the direction and magnitude of an applied \emph{in-plane} magnetic field.
We show how to derive an elegant expression for the field-dependent critical current in a semi-classical limit (where the system is large compared to its Fermi wave length), which allows for straightforward numerical evaluation of the current.
Assuming that we can describe the dynamics of the holes in the normal region with a simple $4\times 4$ Luttinger Hamiltonian and that the carrier density is low enough that only the lowest (heavy-hole) subband is occupied, we identify several different parameter regimes where different spin-mixing mechanisms could be dominating and we calculate the field-dependent critical current in all these regimes.
We are able to connect each mechanism to clear qualitative features in the pattern of critical current that emerges and we present analytical expressions for the current in most limiting cases.
Our results could thus help distinguishing the dominating spin-mixing process at play in an experiment, and as such give insight in the strength and nature of the underlying spin--orbit and Zeeman couplings in the system.

The rest of the paper is organized as follows.
In Sec.~\ref{sec:supercurrent} we will introduce the setup we consider and the model we use to describe it.
We outline our method of calculating the critical current through the junction and explain how we tailor it to the situation where all transport in the normal region is carried by the heavy holes.
In Sec.~\ref{sec:paramlim} we present both our numerical and analytical results, systematically going through the different parameter regimes that could be reached.
Finally, in Sec.~\ref{sec:conclusion} we present a short conclusion.

\begin{figure}[t]
    \centering
    \includegraphics[width = 0.4\textwidth]{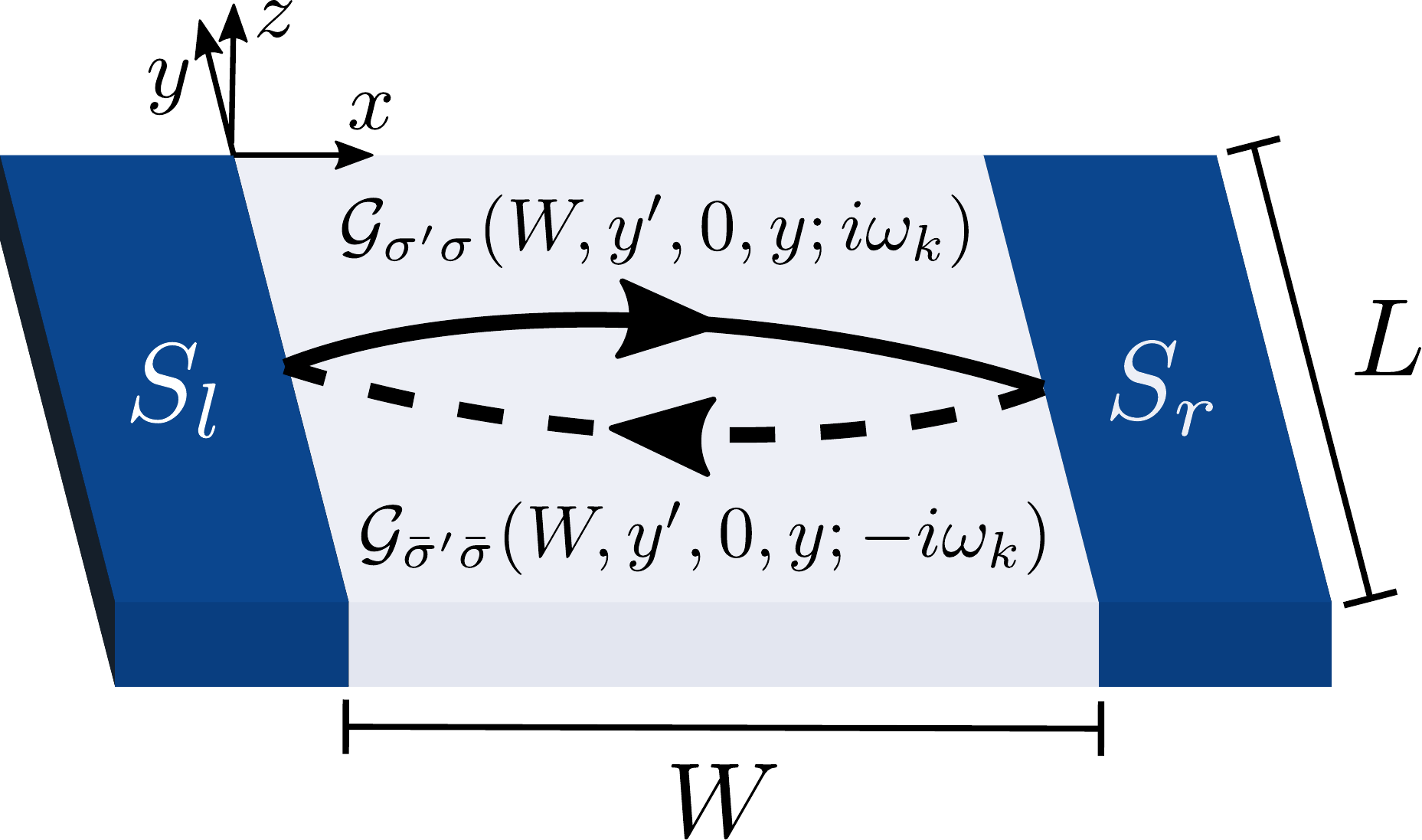}
    \caption{Schematic of the SNS junction: two identical conventional superconductors $S_l$ and $S_r$, connected by a 2D hole gas. The junction has a length of $L$ and a width of $W$, as indicated. An example of a diagram contributing to the Cooper-pair propagator is drawn in the normal region, where the electron(hole) propagator is depicted with a solid(dashed) line.}
    \label{fig:system}
\end{figure}

\section{\label{sec:supercurrent}Model}

Fig.~\ref{fig:system} shows a cartoon of the system we consider:
A 2DHG is contacted from two sides by two identical conventional superconductors to create an SNS junction, where we assume the coupling between the superconductors and the normal region to be weak.
We define the length $L$ and width $W$ of the junction as indicated in the Figure and choose the coordinate system such that the average flow of supercurrent is in the $x$-direction and the out-of-plane direction is denoted by $z$.
We assume the clean junction limit, as the width of experimentally viable devices is typically of the order $100~\text{nm}$ to $1~\text{µm}$ \cite{Hendrickx2019a,Vigneau2019,Aggarwal2021}, while the mean free path of, e.g., Ge 2DHGs has been measured to be up to $\sim 6~\text{µm}$~\cite{Hendrickx2019,Aggarwal2021}.

We will first introduce the method we chose for calculating the supercurrent through the junction.
In the ground state, the current is given by
\begin{equation}
    I(\phi) = \frac{2e}{\hbar} \frac{\partial F}{\partial \phi},
\end{equation}
where $F$ is the free energy of the junction and $\phi $ is the difference in phase between the two superconductors.

We describe the coupling between the hole gas and the superconducting leads with a tunneling Hamiltonian
\begin{equation}
    \begin{split}
        H_t = \sum_\sigma\int \!dy \,\big[ &t_{l}\hat{\psi}_\sigma^\dagger(0,y)\hat{\Psi}_{\sigma,L}(0,y) \\
        &+ t_{r}\hat{\psi}_\sigma^\dagger(W,y)\hat{\Psi}_{\sigma,R}(W,y) + \text{H.c.} \big] ,
    \end{split}
\end{equation}
where $\hat \psi^\dagger_\sigma ({\bf r})$ is the creation operator for an electron with spin $\sigma$ at position ${\bf r}$ in the normal region, and $\hat \Psi^\dagger_{\sigma,L(R)}({\bf r})$ for an electron with spin $\sigma$ at position ${\bf r}$ in the left(right) superconductor.
The lines $x=0,W$ define the interfaces between the superconductors and the normal region.

We assume the coupling amplitudes $t_{l,r}$ to be small enough to justify a perturbative treatment of $H_t$.
Weak coupling can result from, e.g., interfacial disorder, but could also be a consequence of the difference in underlying orbital structure of the electronic wave functions in the superconductors' conduction band and the semiconductor's valence band.
The leading-order correction to $F$ that depends on $\phi$ is second order in the self energy due to the proximity of the superconductors, or fourth order in the coupling Hamiltonian $H_t$~\cite{mahan:book},
\begin{equation}
    F^{(4)} = -\frac{1}{4!\beta} \int_0^\beta \!\! d\tau_{1...4}\langle \hat T_\tau H_t(\tau_1)H_t(\tau_2)H_t(\tau_3)H_t(\tau_4)\rangle,
    \label{eq:Omega4}
\end{equation}
where $\beta = 1/T$ is the inverse temperature, $\hat T_\tau$ is the (imaginary) time-ordering operator, and $\hbar = k_{\rm B}= 1$.
In the evaluation of (\ref{eq:Omega4}) we focus on the fully connected diagrams only, since those are the ones that can probe the phase difference between the two superconductors.

Anticipating that we will make a semi-classical approximation later, assuming that the dimensions of the junction are much larger than the Fermi wave length $\lambda_{\rm F}$, we will take the Andreev reflection at the NS interface to be local and energy-independent.
After applying Wick's theorem to the correlator in (\ref{eq:Omega4}) this allows us to simplify the correction to
\begin{equation}\label{eq:energyCorr}
\begin{split}
   F^{(4)}= -\lambda_l & \lambda_r \iint\! dy\,dy'\\
   & \times\text{Re}\left\{ e^{i[\varphi_l(y) - \varphi_r(y')]} C(W,y^\prime;0,y)\right\},
\end{split}
\end{equation}
where $\lambda_{l,r} = \pi t_{l,r}^2 \nu_{\rm eff}$ parameterize the strength of the coupling to the superconducting leads, with $\nu_{\rm eff}$ the local effective one-dimensional tunneling density of states of the superconductors (giving the $\lambda$'s dimensions energy$\times$meters).
The phase difference
\begin{equation}
    \varphi_{l}(y) - \varphi_r(y') = \phi +  \frac{\pi (y+y') B_{z} W}{\Phi_0},
\end{equation}
with $\Phi_0 = h/2e$ the flux quantum, depends on the two $y$-coordinates in such a way that it captures the coupling to an out-of-plane magnetic field~\cite{Hart2017}, due to the flux $\Phi = B_zWL$ penetrating the junction.
We assume that the magnetic field $B_z$ is small enough that it does not significantly affect the trajectories of the charges~\cite{Note1}. 
We used the function
\begin{equation}
    \begin{split} 
        C({\bf r}';{\bf r})= \frac{T}{2} \sum_{k} {\rm Tr} \big[
        \bar{\cal G}({\bf r}',{\bf r};i\omega_{k})\sigma_y
        \bar{\cal G}({\bf r}',{\bf r};-i\omega_{k})^T\sigma_y \big],
    \end{split}
\end{equation}
were the $\bar {\cal G}$ are 2$\times$2 matrices in spin space,
\begin{equation}
    \bar{\cal G}({\bf r}',{\bf r};i\omega_k) = \left(
    \begin{array}{cc}
    {\cal G}_{\ua\ua}({\bf r}',{\bf r};i\omega_k) & {\cal G}_{\ua\da}({\bf r}',{\bf r};i\omega_k) \\
    {\cal G}_{\da\ua}({\bf r}',{\bf r};i\omega_k) & {\cal G}_{\da\da}({\bf r}',{\bf r};i\omega_k)
    \end{array}
    \right),
\end{equation}
with ${\cal G}_{\sigma'\sigma}({\bf r}',{\bf r};i\omega_{k}) = - \int_0^\beta \! d\tau\, e^{i\omega_k\tau} \langle \hat T_\tau \hat \psi_{\sigma'}({\bf r}',\tau)\hat \psi^\dagger_\sigma({\bf r},0)\rangle$ the thermal Green function at Matsubara frequency $\omega_k = (2k+1)\pi T$ for (spin-dependent) electronic propagation in the normal region.
The correlation function $C(W,y^\prime;0,y)$ as used in (\ref{eq:energyCorr}) can thus be interpreted as the probability amplitude for a Cooper pair to cross the junction, from the point $(0,y)$ to the point $(W, y^\prime)$, as illustrated by the simple diagram shown in Fig.~\ref{fig:system}.
In writing Eq.~(\ref{eq:energyCorr}) we further assumed the pairing in the superconductors to be conventional $s$-type, described by pairing terms like $H_{\rm pair}^{(S)} = -\sum_{{\bf k}} \left\{ \Delta_0 \hat\Psi_{{\bf k}\uparrow}^\dagger \hat\Psi_{-{\bf k}\downarrow}^\dagger +  \Delta_0^* \hat\Psi_{-{\bf k}\downarrow} \hat\Psi_{{\bf k}\uparrow}\right\}$.
Within all approximations made, other details of the dynamics inside the superconductors will only affect the magnitude of the two coupling parameters $\lambda_{l,r}$.

We assume that the carriers in the normal region can be described by a 4$\times$4 Luttinger Hamiltonian~\cite{winkler2003},
\begin{equation}\label{eq:lutt}
    H_0 = \frac{1}{2m_0}
    \begin{pmatrix}
        P+Q     & 0     & 0     & M     \\
        0       & P+Q   & M^*   & 0  \\
        0       & M     & P-Q   & 0     \\
        M^*     & 0     & 0     & P-Q   \\
    \end{pmatrix},
\end{equation}
where
\begin{subequations}
\begin{align}
P &= \gamma_1\left(k^2 + \langle k_z^2 \rangle\right),\\
Q & = \gamma_2\left(k^2 -2\langle k_z^2\rangle \right),\\
M & = -\tfrac{1}{2}\sqrt{3}\left[(\gamma_2+\gamma_3)k_-^2 +(\gamma_2-\gamma_3)k_+^2\right],\label{eq:M}
\end{align}
\end{subequations}
using $k_\pm = k_x \pm ik_y$ and $k^2 = k_x^2+k_y^2$, and with $m_0$ being the bare electron mass and $\gamma_{1,2,3}$ the three dimensionless material-specific Luttinger parameters. 
This Hamiltonian is written in the basis of the angular-momentum states $\{ \ket{+3/2}, \ket{-3/2}, \ket{+1/2}, \ket{-1/2} \}$ with total angular momentum $3/2$ and it includes an extra minus sign, i.e., it describes the dynamics from a hole perspective.
The $z$-coordinate (along which the holes are strongly confined) has already been integrated out, $\langle k_z^2 \rangle \sim 1/d^2$, with $d$ the transverse confinement length, and we neglect the effects of strain for simplicity.

We now assume that transverse confinement is strong enough to make the splitting $\delta_{HL} = 2 \gamma_2 \langle k_z^2 \rangle / m_0$ the largest energy scale involved, on the order of $\sim 100 \text{meV}$ in planar Ge \cite{Hendrickx2018,Sammak2019,Terrazos2021}, which allows us to focus on the so-called heavy-hole (HH) subspace $\{ \ket{+3/2},\ket{-3/2}\}$ and treat the coupling to the light-hole (LH) states $\{ \ket{+1/2},\ket{-1/2}\}$ perturbatively.
We will further assume that the Andreev reflection at the interfaces with the superconductors pairs hole states with opposite orbital and spin angular momentum, such as $\ket{\pm 3/2}$.
This allows us to treat the low-energy HH subspace $\{ \ket{+3/2},\ket{-3/2}\}$ as an effective spin-1/2 system that can host a supercurrent that can be described with the formalism presented above~\cite{Note2}.

\begin{figure}[b!]
    \centering
    \includegraphics[width=0.28\textwidth]{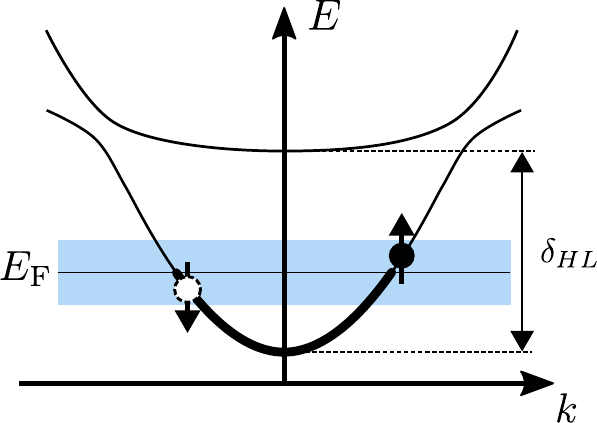}
    \caption{A sketch of the spectrum of the Hamiltonian \eqref{eq:2DLutt}, where we assumed the perturbations $H_{\rm Z}$ and $H_{\rm R}$ to be small enough to be neglected.
    We see the heavy-hole and light-hole bands being split by the HH--LH splitting $\delta_{HL}$ and anticross where they are mixed by the off-diagonal terms $k_\pm^2/2m_x$.
    The blue shaded region around the Fermi energy, $E_{\rm F}$, indicates the energy window within which all relevant dynamics are assumed to happen, its width being of the order of $|\Delta_0|$.
    Superconducting pairing in the 2DHG is induced between holes in the HH band with opposite spin and momentum, as illustrated in the Figure.}
    \label{fig:bandStructure}
\end{figure}

Furthermore, we want to include the Zeeman effect due to an in-plane magnetic field ${\bf B}_\parallel$ as well as Rashba-type spin--orbit coupling.
We describe the in-plane Zeeman effect with the Hamiltonian
\begin{equation}
    H_{\rm Z} = -2\kappa \left(B_+ J_- + B_-J_+\right),
\end{equation}
where $J_{\pm} = J_x \pm iJ_y$ are the spin-3/2 raising and lowering operators, $B_\pm= B_x \pm iB_y$, the hole $g$-factor is $\kappa$, and we set $\mu_B=1$.
The spin--orbit coupling, which can be due to asymmetries in the confining potential or to an externally applied out-of-plane electric field, is described with
\begin{equation}
    H_{\rm R} = i\alpha_{\rm R} \left( k_+ J_- -  k_- J_+\right),
\end{equation}
where $\alpha_{\rm R}$ characterizes the strength of the coupling.

We add these two ingredients to the projected two-dimensional Luttinger Hamiltonian introduced above and we make the so-called spherical approximation, amounting to the assumption $|\gamma_2 - \gamma_3| \ll \gamma_2 + \gamma_3$, which allows to drop the last term in (\ref{eq:M}).
For many commonly used semiconductors, such as Ge, GaAs, InSb, and InAs (but not for Si), this is a valid approximation~\cite{winkler2003}.
Otherwise we impose no constraints on the Luttinger parameters.
Then the total Hamiltonian for the hole gas is
\begin{widetext}
\begin{align}\label{eq:2DLutt}
    H_{\rm tot} & = H_0 + H_{\rm Z} + H_{\rm R} \nonumber\\
    &=\begin{pmatrix}
        k^2/2m_H     & 0                    & -\sqrt{3}\left(2\kappa B_- + i\alpha_{\rm R} k_-\right)     & -k_-^2/2m_x      \\
        0                    & k^2/2m_H     & -k_+^2/2m_x      & \sqrt{3}\left(-2\kappa B_+ + i\alpha_{\rm R} k_+\right)  \\
        \sqrt{3}\left(-2\kappa B_+ + i\alpha_{\rm R} k_+\right)& -k_-^2/2m_x & \delta_{HL} + k^2/2m_L & -4\kappa B_- -4 i\alpha_{\rm R} k_-  \\
        -k_+^2/2m_x & -\sqrt{3}\left(2\kappa B_- + i\alpha_{\rm R} k_-\right)  & -4\kappa  B_+ + 4i\alpha_{\rm R} k_+ & \delta_{HL} + k^2/2m_L \\
    \end{pmatrix},
\end{align}
where we introduced the effective HH and LH masses $m_H = m_0/(\gamma_1+ \gamma_2)$ and $m_L = m_0/(\gamma_1- \gamma_2)$.
We further used $m_x =2m_0 /\sqrt{3}(\gamma_2+\gamma_3)$, which governs the strength of the momentum-dependent HH-LH mixing. A scetch of Hamiltonian \eqref{eq:2DLutt} can bee seen in Fig. \ref{fig:bandStructure}.

Our assumption that $\delta_{HL}$ is the largest energy scale involved allows us to treat the HH--LH coupling perturbatively.
We first diagonalize the LH subspace in (\ref{eq:2DLutt}), after which we perform a Schrieffer-Wolff transformation to decouple the HH and LH subspaces.
To second order in $1/\delta_{HL}$ we find the effective HH Hamiltonian
\begin{equation}\label{eq:hheff}
    H_{HH} = \left( \begin{array}{cc}
    k^2/2m_H & -\delta_{HL}^{-1}(k_-^2/m_x) \beta_1 - 4\delta_{HL}^{-2}\beta_1^2\beta_2 \\
    -\delta_{HL}^{-1}(k_+^2/m_x) \beta_1^* - 4\delta_{HL}^{-2} (\beta_1^*)^2\beta_2^* & k^2/2m_H
    \end{array} \right),
\end{equation}
\end{widetext}
where we ignored the shift of the diagonal elements, and we used $\beta_1 = \sqrt{3}(2\kappa B_- + i\alpha_{\rm R} k_-)$ and $\beta_2 = \kappa B_- + i\alpha_{\rm R} k_-$.
We see that, depending on the magnitude of $\delta_{HL}$, the typical in-plane (Fermi) momentum $k_{\rm F}$ of the current-carrying holes, the strength of the spin--orbit coupling $\alpha_{\rm R}$, and the magnitude of the applied in-plane magnetic field, different terms can dominate the effective coupling of the two HH states.
We used that in the perturbative limit we consider here one always has $(\delta_{HL}^{-1} k_\pm^2 / m_x)^2 \ll \delta_{HL}^{-1} k_\pm^2 / m_x$, and we thus ignore the contribution $-4\delta_{HL}^{-2}(k_-^2/m_x)^2(\beta_2^*\sigma_+ + \beta_2 \sigma_-)$ to Eq.~(\ref{eq:hheff}).

We can now consider different cases.
Firstly, for a very thin 2DHG we can assume that the term $\propto \delta_{HL}^{-1}$ will dominate, which leaves two qualitatively different coupling terms in $H_{HH}$,
\begin{align}
     H_{0,3}^{(1)} = & \frac{-i\sqrt 3 \alpha_{\rm R}}{m_x\delta_{HL}}\left(k_-^3 \sigma_+ - k_+^3 \sigma_-\right),\label{eq:coup03}\\
     H_{1,2}^{(1)} = &  \frac{-2\sqrt 3 \kappa}{m_x\delta_{HL}} \left( B_-k_-^2 \sigma_+ + B_+ k_+^2 \sigma_-\right),\label{eq:coup12}
\end{align}
where the subscripts of $H$ refer to the powers of $B$ and $k$ appearing in the term, respectively, and the superscript indicates the power of $\delta_{HL}^{-1}$.
If the 2DHG is less thin, then the term $\propto \delta_{HL}^{-2}$ in (\ref{eq:hheff}) could also contribute, which allows for four additional coupling terms,
\begin{align}
     H_{0,3}^{(2)} {} & {} = 12\frac{i\alpha^3_{\rm R}}{\delta^2_{HL}}\left(k_-^3 \sigma_+ - k_+^3 \sigma_-\right),\label{eq:coup03b} \\
     H_{1,2}^{(2)} {} & {} =   60\frac{\alpha_{\rm R}^2 \kappa}{\delta^2_{HL}} \left( B_-k_-^2 \sigma_+ + B_+ k_+^2 \sigma_-\right), \label{eq:coup12b}\\
     H^{(2)}_{2,1} & = -96\frac{i\alpha_{\rm R}\kappa^2}{\delta^2_{HL}} \left( B_-^2k_- \sigma_+ - B_+^2k_+ \sigma_-\right), \\
    H^{(2)}_{3,0} & = -48\frac{\kappa^3}{\delta^2_{HL}} \left( B_-^3 \sigma_+ + B_+^3 \sigma_-\right).\label{eq:coup30}
\end{align}

We now make the assumption that all relevant dynamics happen on an energy scale very close to the Fermi level $E_{\rm F}$.
This allows us to linearize the kinetic energy in $H_{HH}$ in ${\bf k}$ and to assume that the magnitude of the in-plane momentum $k \approx k_{\rm F}$ in the off-diagonal terms.
This leaves us with a general $2\times 2$ Hamiltonian effectively describing the HH subsystem (up to a constant)
\begin{equation}
    H_{HH} = v_{\rm F} (k-k_{\rm F}) + \boldsymbol{\beta}(\theta)\cdot\boldsymbol{\sigma},\label{eq:HHH}
\end{equation}
where $v_{\rm F} = k_{\rm F}/m_H$ is the Fermi velocity and the field $\boldsymbol\beta$ includes the off-diagonal terms of $H_{HH}$, depending only on the angle $\theta$, the in-plane direction of ${\bf k}$.
The vector $\boldsymbol \sigma = \{\sigma_x, \sigma_y, \sigma_z\}$ consists of the three Pauli matrices.

Following the approach of Ref.~\cite{Hart2017}, we recognize that $H_{HH}$ in (\ref{eq:HHH}) can be diagonalized in spin space and we denote the two ${\bf k}$-dependent eigenspinors with $\ket{\lambda_{\bf k}}$, where $\lambda = \pm$.
This allows to rewrite Eq.~(\ref{eq:HHH}) as
\begin{equation}
    H_{HH} = \sum_{\lambda_{\bf k} = \pm_{\bf k}} \epsilon_{{\bf k}\lambda} P^{\lambda_{\bf k}},
\end{equation}
in terms of the energies $\epsilon_{{\bf k}\lambda} = v_{\rm F}(k-k_{\rm F}) + \lambda |\boldsymbol \beta(\theta)|$ and the projectors $P^{\lambda_{\bf k}} = \ket{\lambda_{\bf k}}\bra{\lambda_{\bf k}} = \frac{1}{2}[1 + \lambda \hat{\boldsymbol\beta} (\theta)\cdot \boldsymbol\sigma]$, where the dimensionless vector $\hat{\boldsymbol\beta}(\theta) = {\boldsymbol\beta}(\theta)/|{\boldsymbol\beta}(\theta)|$ points along the direction of the field ${\boldsymbol\beta}(\theta)$.

Assuming for simplicity translational invariance inside the 2DHG, the correlation function $C({\bf r}';{\bf r})$ is only a function of the difference in coordinates and reduces at zero temperature to (see the Supplementary Material for a more detailed derivation)
\begin{align}
    C({\bf r}) = \int\!\!\!\int_0^\infty \frac{d\epsilon\,d\epsilon'}{2(\epsilon+ \epsilon')}
    {\rm Tr}\big[ {} & {} \bar g({\bf r},-\epsilon) \sigma_y \bar g({\bf r},-\epsilon')^T \sigma_y\nonumber \\
    {} & {} + \bar g({\bf r},\epsilon) \sigma_y \bar g({\bf r},\epsilon')^T \sigma_y  \big],
\end{align}
using the propagator
\begin{equation}
    \bar g({\bf r},\epsilon) = 
    \frac{1}{(2\pi)^2}\sum_{\lambda_{\bf k} = \pm_{\bf k}} \int d{\bf k} \, e^{i{\bf k}\cdot{\bf r}} \delta( \epsilon - \epsilon_{{\bf k}\lambda})  P^{\lambda_{\bf k}}.
\end{equation}

We then additionally assume that $k_{\rm F}r \gg 1$ for all distances $r = |{\bf r}|$ of interest (such as $W$), which amounts to employing a semi-classical approximation.
In that case one finds that the only momenta that contribute significantly in the propagators $\bar g({\bf r},\epsilon)$ have a wave vector ${\bf k}$ parallel or anti-parallel to ${\bf r}$ (see the Supplementary Material for a formal derivation).
In this limit we find a greatly simplified approximation for the Cooper-pair propagator,
\begin{align}\label{eq:cooperpairprop}
    C({\bf r}) ={} & {}\frac{K}{r^2}\bigg\{
     \cos\left(\frac{|\boldsymbol\beta(\theta)|r}{v_{\rm F}}\right) \cos\left(\frac{|\boldsymbol\beta(\bar\theta)|r}{v_{\rm F}}\right) \\
    {} & {} - \hat{\boldsymbol\beta}(\theta)\cdot\hat{\boldsymbol\beta}(\bar\theta) \sin\left(\frac{|\boldsymbol\beta(\theta)|r}{v_{\rm F}}\right) \sin\left(\frac{|\boldsymbol\beta(\bar\theta)|r}{v_{\rm F}}\right)\bigg\},\nonumber
\end{align}
where the constant $K = k_{\rm F}/(2\pi)^2v_{\rm F}$, and $\theta$($\bar\theta$) is now the in-plane direction parallel(anti-parallel) to ${\bf r}$, which means ${\bf r} = \{r \cos\theta, r\sin\theta\}$ and $\bar\theta = \theta-\pi$.

We note that by inserting this propagator in Eq.~(\ref{eq:energyCorr}) to evaluate the supercurrent we only account for the contribution of straight trajectories between the two superconducting contacts, i.e., we neglect trajectories that involve scattering off the edges of the 2DHG.
This approximation becomes better with increasing aspect ratio $L/W$ of the junction.

\section{Results}\label{sec:paramlim}

We now have all ingredients needed to calculate the supercurrent through the junction, to leading order in the SN coupling strength and within a semi-classical approximation.
In this Section we will present our results.

Due to the large number of competing coupling terms we consider (\ref{eq:coup03}--\ref{eq:coup30}), the field $\boldsymbol\beta(\theta)$, and thus the current, can look very different depending on the parameters one assumes.
However, in the limiting case where only one of the six terms dominates, the Cooper-pair propagator immediately simplifies further:
Assuming that one term $H_{n,m}$ is by far the largest (where, again, $n$ and $m$ refer to the powers of $B$ and $k$ in the coupling term), one finds
\begin{equation}\label{eq:beta}
    \hat{\boldsymbol\beta}(\theta)\cdot\hat{\boldsymbol\beta}(\bar\theta) = \cos(m \pi),
\end{equation}
and in that case the expression given in (\ref{eq:cooperpairprop}) reduces to
\begin{align}
    C({\bf r}) = \frac{K}{r^2} \times \begin{cases}
     \cos\left(2r |\boldsymbol\beta|/v_{\rm F}\right)  & \text{for } m \text{ even} \\
  1
      & \text{for } m \text{ odd}
\end{cases},
\end{align}
where we used that all coupling terms listed above correspond to fields for which $|\boldsymbol\beta(\theta)|$ is independent of $\theta$.

\begin{figure}[]
    \centering
    \includegraphics[width=0.5\textwidth]{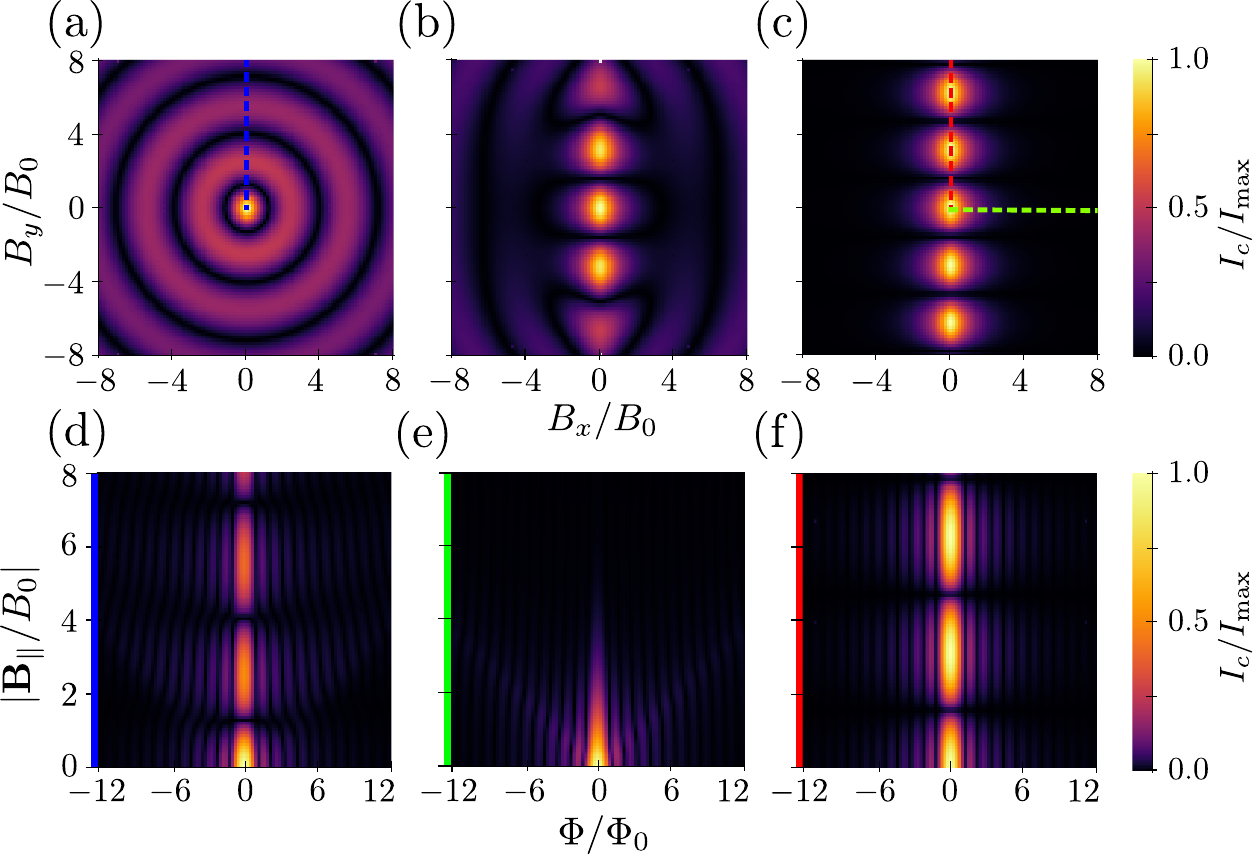}
    \caption{(a--c) Numerically calculated critical current as a function of in plane magnetic field, $B_\parallel = (B_x,B_y)$ and $B_z = 0$, expressed in units of $B_0 = \delta_{HL}v_{\rm F}/8\sqrt 3 \kappa E_x W$.
    We work in the limit of large HH--LH splitting everywhere, using a field $\boldsymbol \beta (\theta)$ as defined by (\ref{eq:couplingFieldLargeHHLH}).
    We show results for different values of $E_{so}$:
    (a) $E_{so}/\kappa B_0=0.2$ (b) $E_{so}/\kappa B_0=2$, and (c) $E_{so}/\kappa B_0=20$.
    (d--f) Critical current as a function of in-plane magnetic field (vertical axis) along given directions, and an additional out-of-plane magnetic field (horizontal axis), quantified by the total flux trough the junction $\Phi$. 
    The colors at the vertical axes correspond to the colored dashed lines in (a--c), which indicate the direction of the in-plane field.
    For all plots the aspect ratio is set at $L/W = 10$, the scale of the critical current is described by $I_{\text{max}} = I_0 \pi L/W$, and $\Phi_0 = h/2e$ is the flux quantum.}
    \label{fig:2DEGEQ}
\end{figure}

In this case the supercurrent through the junction will thus not depend on the direction of the applied in-plane field, only on its magnitude if the coupling term has an even power of $k$.
This is indeed what one expects: 
When the field $\boldsymbol \beta$ is even in momentum, a pairing of opposite spins at the Fermi level introduces a finite average Cooper-pair momentum which is to first approximation linear in the magnitude of the field.
For fields that are odd in momentum, the sign change of $\boldsymbol\beta$ upon inversion of ${\bf k}$ guarantees that there are always eigenstates with opposite spin and momentum available at the Fermi level, independent of the magnitude of the total field.

A more interesting dependence on the magnitude and direction of the in-plane field can arise when two or more coupling terms with different dependence on $B$ and $k$ compete~\cite{Hart2017}.
Calculating the supercurrent numerically for an arbitrary combination of coupling mechanisms is straightforward.
However, to structure our discussion and to gain qualitative insight in the significance of all terms, we will mostly consider limiting cases below, where only a few terms play a role.

\subsection{Large HH--LH splitting}\label{sec:largeHH-LH}

The first case we will investigate is when we have a large HH--LH splitting $\delta_{HL}$ (corresponding to tight out-of-plane confinement).
In that case, the terms (\ref{eq:coup03b}--\ref{eq:coup30}), which are proportional to $\delta_{HL}^{-2}$, are suppressed and the dominating coupling terms are $H_{0,3}^{(1)}$ and $H_{1,2}^{(1)}$.
More quantitatively, we see that this regime is reached when
\begin{gather}\label{eq:ineqlargedelta}
    \sqrt{\delta_{HL} E_x} \gg E_{\rm Z},\, E_{so},
\end{gather}
where we introduced the Zeeman energy $E_{\rm Z} = \kappa B_\parallel$, the spin--orbit energy $E_{so} = \alpha_{\rm R} k_{\rm F}$, and the orbital coupling energy $E_x = k^2_{\rm F}/2m_x$.
(Assuming that the Luttinger parameters are of order unity \cite{Note3}, this last energy scale is of the order of the Fermi energy in the valence band and can thus be tuned by varying the carrier density.)
In this case, the total coupling field is defined by
\begin{align}\label{eq:couplingFieldLargeHHLH}
    \beta_+(\theta) = \frac{2\sqrt{3}E_x}{\delta_{HL}}\left( iE_{so} e^{3i\theta} - 2E_{\rm Z}e^{i(2\theta+\phi_B)} \right),
\end{align}
where $\phi_B$ is the direction of the in-plane magnetic field.

As mentioned, this effective field allows us to calculate the supercurrent and hence the critical current through the junction.
In Fig.~\ref{fig:2DEGEQ} we show the dependence of the resulting critical current on the applied magnetic field, for different ratios of $E_{so}/E_{\rm Z}$:
In the top row of panels we plot the critical current as a function of the the magnitude and direction of a purely in-plane field.
In plots (a--c) we used $E_{so}/\kappa B_0=0.2$, $2$, and $20$, respectively, where $\kappa B_0 = \delta_{HL}v_{\rm F}/8\sqrt 3 E_x W$ and we used an aspect ratio of $L/W = 10$.
For reference, and to compare with Ref.~\cite{Hart2017}, we show in the bottom row the Fraunhofer-like patterns of critical current that emerge when, in addition to an in-plane field (vertical axes), a small perpendicular magnetic field $B_z$ is applied (horizontal axes).
The direction of each ${\bf B}_\parallel$ is indicated with a dashed line in the plots in the top row:
In Fig.~\ref{fig:2DEGEQ}(d) we have $E_{so}/\kappa B_0 =0.2$, as in (a), while the in-plane field is oriented along $\hat y$ (blue dashed line).
In Figs.~\ref{fig:2DEGEQ}(e,f) we used $E_{so}/\kappa B_0=20$ with (e) the in-plane field along $\hat x$ [green dashed line in (c)] and (f) the field along $\hat y$ [red dashed line in (c)].

We see that these results are similar to those of Ref.~\cite{Hart2017} for the case where the competition between Zeeman and Rashba coupling is investigated (cf.~Fig.~3 in Ref.~\cite{Hart2017}).
This can be easily understood from the structure of the semi-classical Cooper-pair propagator (\ref{eq:cooperpairprop}), which only depends on the relative orientation and the magnitude of the fields acting on the ``electrons'' and ``holes'' that propagate with opposite momentum.
The main difference in our coupling terms as compared to the electronic case studied in Ref.~\cite{Hart2017} is an extra factor $k_\pm^2$ due to the intrinsic HH--LH mixing in the valence band.
This additional factor only serves to rotate all effective fields by the same amount, thereby not affecting the direction-dependence of the Cooper-pair propagator.

As expected, in the limit of dominating Zeeman coupling [Fig.~\ref{fig:2DEGEQ}(a)] the period of the oscillations is independent of the direction of propagation of the Cooper pair, since spin rotations are in this case not related to the direction of propagation of Cooper pairs.
In the spin--orbit-dominated case the oscillations in $C({\bf r})$ occur always in a direction perpendicular to ${\bf B}_\parallel$ for the case of the Rashba-type spin--orbit coupling assumed here.

As was pointed out in Ref.~\cite{Hart2017}, in the two limiting cases of strongly dominating $E_{so}$ or $E_{\rm Z}$ the propagator simplifies considerably
\begin{align}\label{eq:cp1}
    C({\bf r}) \approx \frac{K}{r^2} \times \begin{cases}
     \displaystyle
     \cos\left(|{\bf d}| r \right)  & \text{for } E_{\rm Z} \gg E_{so} \\
     \displaystyle
     \cos\left([\hat z \times {\bf d}]\cdot {\bf r} \right)  & \text{for } E_{so} \gg E_{\rm Z}
\end{cases},
\end{align}
with ${\bf d} = {\bf B}_\parallel/B_0 W$ being a vector that characterizes the spatial oscillations of the propagator.

The dependence of the supercurrent on the in-plane field as plotted in Fig.~\ref{fig:2DEGEQ}(a--c) follows from evaluating the following integral (setting the electron charge $e=1$),
\begin{equation}
    \label{eq:analyticCritCur}
        I_c({\bf B}_\parallel) = 4 \bigg|\lambda_l\lambda_r \iint_0^L dy\,dy'\, C(W,y'-y) \bigg|,
\end{equation}
which can be performed (semi-)analytically in the two limits discussed above.
For the case of $E_{\rm Z}\gg E_{so}$ we rewrite (\ref{eq:analyticCritCur}) as
\begin{equation}
    I_c({\bf B}_\parallel) = 2 I_0 \int_1^q d\rho \, \frac{\cos (\alpha \rho)}{\rho}
    \left( \frac{L}{W\sqrt{\rho^2-1}} - 1 \right),\label{eq:intappr1}
\end{equation}
where $I_0 = 4K|\lambda_l\lambda_r|$ sets the scale of the supercurrent, we introduced the parameter $q = \sqrt{1+(L/W)^2}$, and we used $\alpha = B_\parallel/B_0$.
In the limit of a long junction, i.e., $L \gg W$, we can approximate $q \to \infty$, yielding
\begin{align}
        I_{c1}({\bf B}_\parallel) = {} & {}
        I_0 \frac{\pi L}{2W}\, \bigg| \pi\alpha \big[J_0(\alpha)H_1(\alpha)-J_1(\alpha)H_0(\alpha)\big]
        \nonumber\\
        {} & {} \hspace{4em} + 2\big[ 1-\alpha J_0(\alpha)\big] \bigg|,
    \label{eq:absValIntegral}
\end{align}
where $J_n(x)$ are Bessel functions of the first kind and $H_n(x)$ are Struve functions.
We note that we neglected the second term in (\ref{eq:intappr1}), which implies the assumption that $\alpha$ is not exponentially small, i.e., $\alpha \gg e^{-L/W}$.
An approximate analytic solution of (\ref{eq:intappr1}) for general $L/W$ is presented in the Supplementary Material.
For the second case, $E_{so} \gg E_{\rm Z}$, we find in the same limit of $L\gg W$
\begin{align}
        I_{c2}({\bf B}_\parallel) \approx {} & {}
        I_0 \frac{\pi L}{W}\, e^{-|B_x/B_0|}|\cos(B_y/B_0)|.
    \label{eq:dotProdIntegral}
\end{align}
Comparing to Fig.~\ref{fig:2DEGEQ}(a,c) we see that the analytic expressions (\ref{eq:absValIntegral},\ref{eq:dotProdIntegral}) indeed capture the behavior of the critical current as a function of ${\bf B}_\parallel$ in the two limiting cases:
For large Zeeman fields the current shows ``damped'' oscillations as a function of $B_\parallel$ and for dominating spin--orbit coupling the behavior becomes direction-dependent, showing oscillations for ${\bf B}_\parallel$ being perpendicular to the mean direction of current flow and rapid decay for ${\bf B}_\parallel$ parallel to the current.

\subsection{Smaller HH--LH splitting}\label{sec:smallHHLH}

The second situation we will consider is when we have a smaller HH--LH splitting and/or orbital coupling $E_x$.
In this case the second-order terms (\ref{eq:coup03b}--\ref{eq:coup30}), which are proportional to $\delta_{HL}^2$, can be dominating.
Formally, this will be the case when
\begin{equation}\label{eq:ineqsmalldelta}
    E_{\rm Z},\,E_{so} \gg \sqrt{\delta_{HL} E_x},
\end{equation}
and the total coupling field then becomes
\begin{equation}\label{eq:couplingFieldSmallHHLH}
    \begin{split}
        \beta_+(\theta) = \frac{12}{\delta_{HL}^2}\Big(&-iE_{so}^3 e^{3i\theta} +5E_{so}^2E_{\rm Z} e^{i(2\theta+\phi_B)} \\
         &+8iE_{so}E_{\rm Z}^2 e^{i(\theta+2\phi_B)} - 4E_{\rm Z}^3e^{3i\phi_B} \Big).
    \end{split}
\end{equation}
If we do not assume anything about the ratio $E_{so}/E_{\rm Z}$ then all four terms could contribute significantly.

\begin{figure}[]
    \centering
    \includegraphics[width=0.5\textwidth]{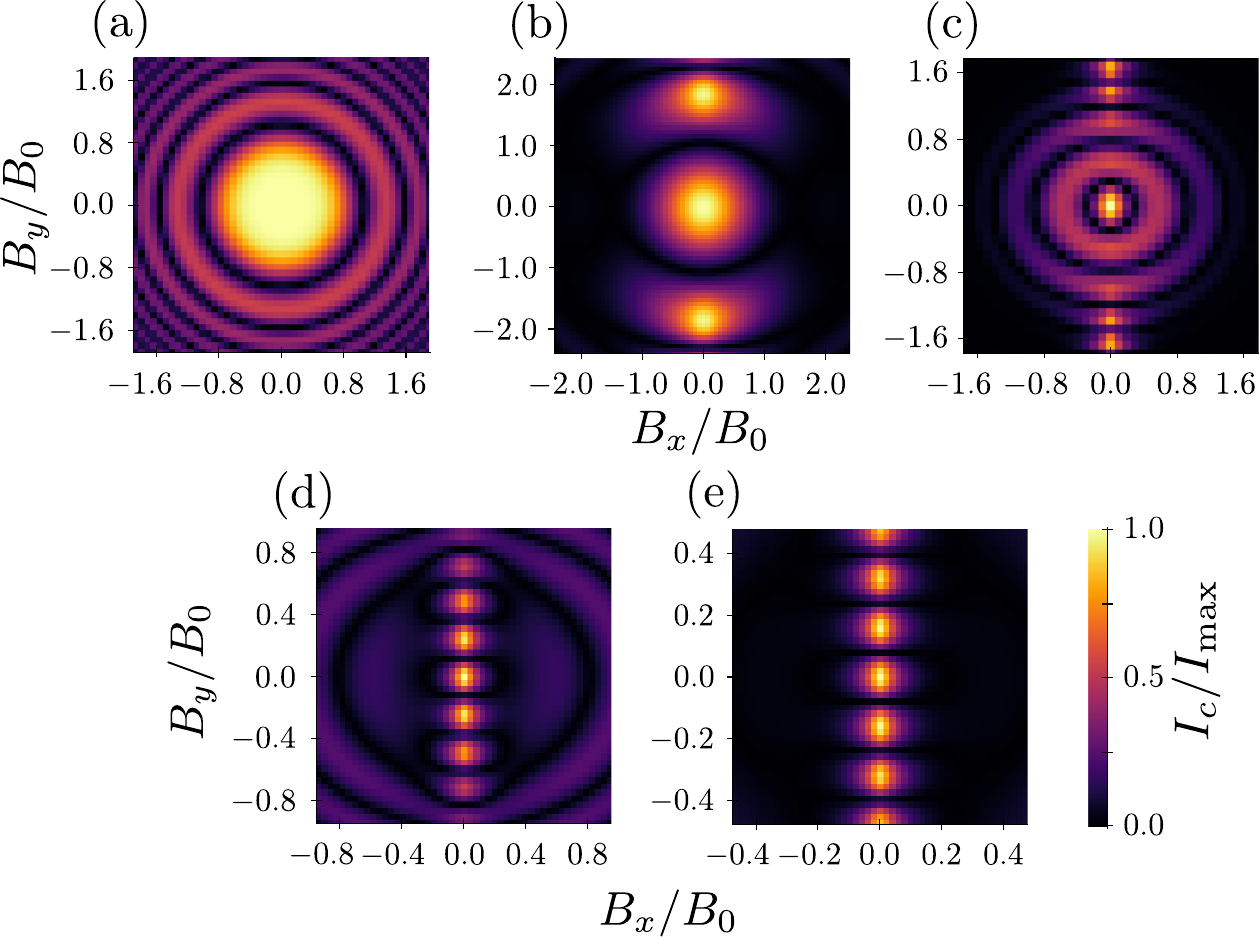}
    \caption{Numerically calculated critical current as a function of in-plane magnetic field, $B_\parallel = (B_x,B_y)$, assuming small HH--LH splitting $\delta_{HL}$.
    We explore (a) $E_{so}/\kappa B_0 = 0.157$, (b) $E_{so}/\kappa B_0 = 1.19$, (c) $E_{so}/\kappa B_0 = 1.97$, (d) $E_{so}/\kappa B_0 = 3.16$, and (e) $E_{so}/\kappa B_0 = 3.94$, where now $\kappa B_0 = (\delta_{HL}^2v_{\rm F}/96 W)^{1/3}$.
    For all plots the aspect ratio is set at $L/W = 10$, the scale of the critical current is described by $I_{\text{max}} = I_0\pi L/W$.}
    \label{fig:2DEGdist}
\end{figure}

We thus assume throughout this Section that the inequality (\ref{eq:ineqsmalldelta}) holds so that the total field $\boldsymbol \beta$ can be approximated by (\ref{eq:couplingFieldSmallHHLH}), and we start by numerically exploring the behavior of the critical current as a function of in-plane field, over a range of $E_{so}/E_{\rm Z}$.
In Fig.~\ref{fig:2DEGdist} we show the calculated critical current, where we used (a) $E_{so}/\kappa B_0 = 0.157$, (b) $E_{so}/\kappa B_0 = 1.19$, (c) $E_{so}/\kappa B_0 = 1.97$, (d) $E_{so}/\kappa B_0 = 3.16$, and (e) $E_{so}/\kappa B_0 = 3.94$, where now $\kappa B_0 = (\delta_{HL}^2v_{\rm F}/96 W)^{1/3}$ and in all cases we again used an aspect ratio of $L/W = 10$.
In the limits of small or large $E_{so}/\kappa B_0$ [Figs.~\ref{fig:2DEGdist}(a,e)] we see qualitatively similar behavior as for the case of large HH--LH splitting [cf.~Fig.~\ref{fig:2DEGEQ}(a,c)], whereas the intermediate regime shows several new features.
With this in mind we now discuss the different parameter regimes, which will provide some understanding of the critical-current patterns we observe.

We first consider the case where the $g$-factor $\kappa$ is relatively large, so that for most fields of interest one has
\begin{equation}\label{eq:largeGInEq}
    E_{\rm Z}  \gg E_{so}  \gg \sqrt{\delta_{HL} E_x}.
\end{equation}
In that case we can approximate
\begin{align}\label{eq:couplingFieldSmallHHLHlargeG}
    \beta_+(\theta) \approx \frac{48 E_{\rm Z}^2}{\delta_{HL}^2}\left( 2iE_{so} e^{i(\theta+2\phi_B)} - E_{\rm Z}e^{3i\phi_B} \right),
\end{align}
i.e., the behavior of $\boldsymbol\beta$ is dominated by the two terms $H^{(2)}_{2,1}$ and $H^{(2)}_{3,0}$.
Comparing this expression with Eq.~(\ref{eq:couplingFieldLargeHHLH}) we see that the effective coupling field is similar to that in the case of large $\delta_{HL}$, the main difference being an additional factor that is quadratic in $E_{\rm Z}$.
This means that, within the range of validity of (\ref{eq:couplingFieldSmallHHLHlargeG}), limiting expressions for the Cooper-pair propagator can be derived that look very similar to Eq.~(\ref{eq:cp1}).
We thus find
\begin{align}
    C({\bf r}) \approx \frac{K}{r^2}\cos\left(|{\bf d}| r \right)  \quad \text{for } E_{\rm Z} \gg E_{so},
\end{align}
where now ${\bf d} = (B_\parallel/B_0)^3({\hat{b}_\parallel}/W)$, with ${\hat{b}_\parallel}$ being the unit vector pointing in the direction of ${\bf B}_\parallel$ and $B_0$ still being defined by $\kappa B_0 = (\delta_{HL}^2v_{\rm F}/96 W)^{1/3}$.
The only difference with the corresponding limit in Eq.~(\ref{eq:couplingFieldLargeHHLH}) is indeed a cubic versus linear dependence on $B_\parallel$, which explains the main difference in appearance between Figs.~\ref{fig:2DEGdist}(a) and \ref{fig:2DEGEQ}(a).
In the limit of a large aspect ratio $L \gg W$ we thus obtain the same approximate analytic expression $I_{c1}({\bf B}_\parallel)$ as presented in Eq.~(\ref{eq:absValIntegral}), the only difference being that one now needs to insert $\alpha = (B_\parallel/B_0)^3$.

When $E_{so}$ is increased, as is done in Fig.~\ref{fig:2DEGdist}(b--e), the opposite limit of $E_{so} \gg E_{\rm Z}$ will of course not be reached while still satisfying (\ref{eq:largeGInEq}).
However, in that hypothetical limit one would find from (\ref{eq:couplingFieldSmallHHLHlargeG}) that $C({\bf r}) \approx (K/r^2) \cos\left([\hat z \times {\bf d}]\cdot {\bf r} \right)$, analogously to the case of large $\delta_{HL}$, again with ${\bf d} = (B_\parallel/B_0)^3({\hat{b}_\parallel}/W)$.
This would ultimately yield the same approximate expression for the critical current in the limit of $L \gg W$ as before,
\begin{align}
        I_{c2}({\bf B}_\parallel) \approx {} & {}
        I_0 \frac{\pi L}{W}\, e^{-\left|  B_\parallel^2B_x/B_0^3\right|}\left|\cos\left( B_\parallel^2B_y/B_0^3 \right)\right|,
    \label{eq:dotProdIntegral2}
\end{align}
again with a cubic instead of linear dependence on the fields.
Although this limit will obviously never be reached, the change in behavior of the critical current from Fig.~\ref{fig:2DEGdist}(a) to (b) can be understood as a first step into the intermediate regime between the two limits, similar to the difference between Figs.~\ref{fig:2DEGEQ}(a) and (b) but now with a cubic dependence on the in-plane field.

We now turn our attention to the opposite case of a relatively small $g$-factor $\kappa$, so that
\begin{equation}\label{eq:smallGInEq}
    E_{so}  \gg E_{\rm Z}  \gg \sqrt{\delta_{HL} E_x}.
\end{equation}
for most fields of interest.
In that case the main competing coupling terms will be $H_{0,3}^{(2)}$ and $H_{1,2}^{(2)}$, yielding approximately
\begin{equation}\label{eq:couplingFieldSmallHHLHLargeRashba}
        \beta_+(\theta) \approx \frac{12E_{so}^2}{\delta_{HL}^2}\Big( - iE_{so} e^{3i\theta} +5E_{\rm Z} e^{i(2\theta+\phi_B)} \Big).
\end{equation}
This is qualitatively the same as the coupling field in (\ref{eq:couplingFieldLargeHHLH}), where the energy scale $E_x$ is replaced by $E_{so}^2/\delta_{HL}$.
The Cooper pair propagator thus becomes
\begin{align}
    C({\bf r}) \approx \frac{K}{r^2}\cos\left([\hat z \times {\bf d}]\cdot {\bf r} \right)  \quad \text{for } E_{so} \gg E_{\rm Z},
\end{align}
with ${\bf d} = (5E_{so}^2/4\kappa^2 B_0^2)(B_\parallel/B_0W)$, still using the same $\kappa B_0 = (\delta_{HL}^2v_{\rm F}/96 W)^{1/3}$, and in the limit $L\gg W$ the critical current takes again the form
\begin{align}
        I_{c2}({\bf B}_\parallel) \approx {} & {}
        I_0 \frac{\pi L}{W}\, e^{-\gamma|B_x/B_0|}|\cos(\gamma B_y/B_0)|,
    \label{eq:dotProdIntegral3}
\end{align}
with $\gamma = 5E_{so}^2/4\kappa^2 B_0^2$.
We indeed see that the numerical results presented in Figs.~\ref{fig:2DEGdist}(e) and \ref{fig:2DEGEQ}(c) coincide, up to scaling factors.

We can again qualitatively understand the phenomenology of the change in the pattern of critical current when moving toward the intermediate regime by decreasing $E_{so}$:
When the Zeeman term becomes more important, the competition between the two terms in (\ref{eq:couplingFieldSmallHHLHLargeRashba}) will start a transition from a periodic critical current pattern along $B_y$ [Fig.~\ref{fig:2DEGdist}(e) and Fig.~\ref{fig:2DEGEQ}(c)] toward a circularly symmetric pattern with a linear dependence on $B_\parallel$ as described by (\ref{eq:absValIntegral}) with $\alpha = \gamma B_\parallel/B_0$ [compare Fig.~\ref{fig:2DEGdist}(d) with Fig.~\ref{fig:2DEGEQ}(b)].
The true limit yielding such a circular pattern will again not be reached since (\ref{eq:couplingFieldSmallHHLHLargeRashba}) will break down already when $E_{so} \lesssim E_{\rm Z}$.

The remaining ``intermediate'' plot shown in Fig.~\ref{fig:2DEGdist}(c) can be roughly interpreted as a hybrid result between the two regimes discussed above:
At small fields, a circularly symmetric linear-in-field limiting pattern emerges that is expected from using (\ref{eq:couplingFieldSmallHHLHLargeRashba}) for small $E_{so}$ (the large-field limit for the case of dominating spin--orbit interaction), which transitions at larger fields into the oscillating pattern along $B_y$ described by (\ref{eq:dotProdIntegral2}) resulting from assuming small $E_{\rm Z}$ in (\ref{eq:couplingFieldSmallHHLHlargeG}) (the small-field limit for dominating Zeeman coupling).

\subsection{Weak spin--orbit coupling}\label{sec:noSOI}

\begin{figure}[t]
    \centering
    \includegraphics[width=0.5\textwidth]{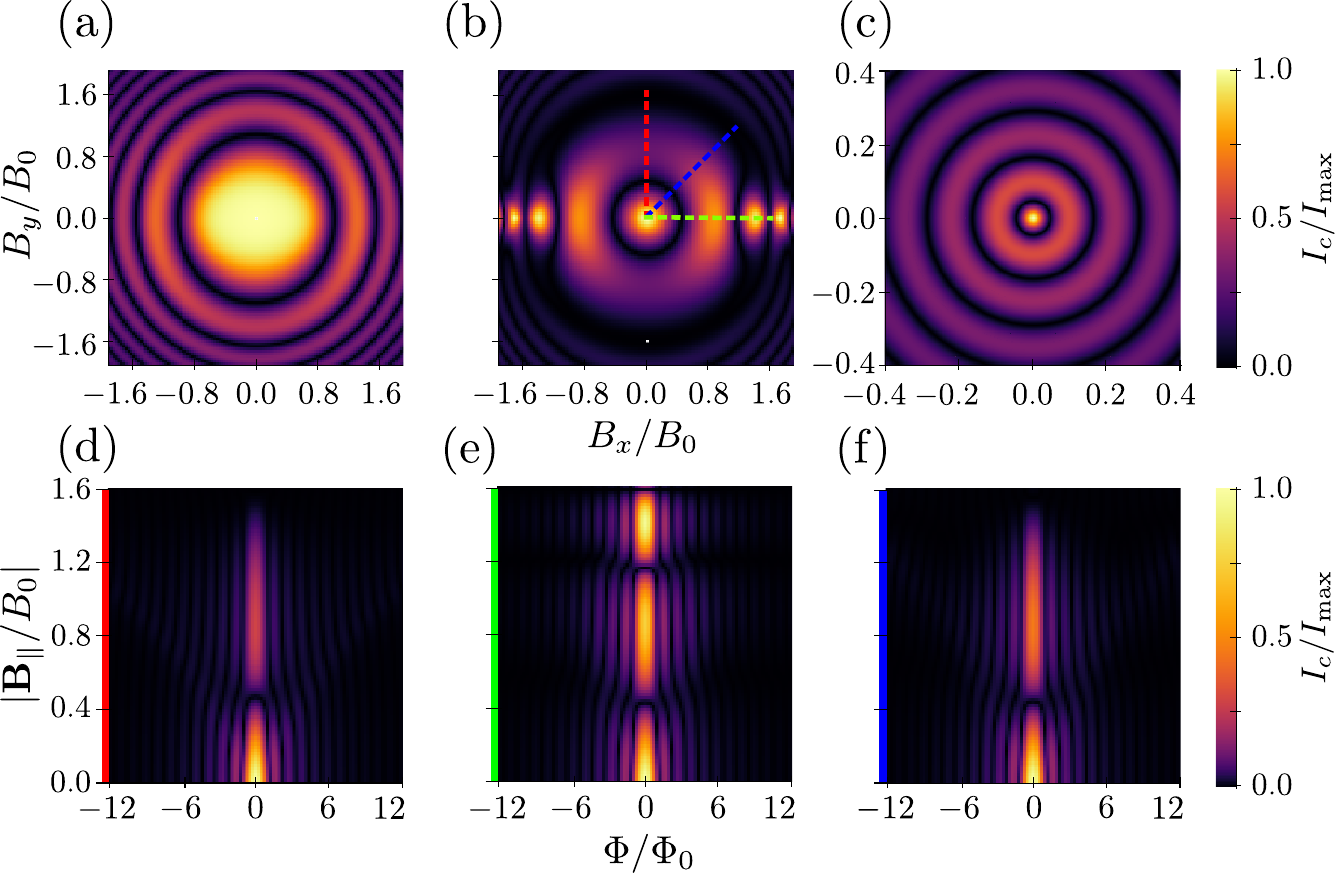}
    \caption{Numerically calculated critical current in the absence of spin--orbit coupling.
    (a--c) The dependence of the current on an in-plane field, where (a) $\sqrt{3}E_{x}\delta_{HL}/12 (\kappa B_0)^2 = 0.252$, (b)  $\sqrt{3}E_{x}\delta_{HL}/12 (\kappa B_0)^2= 2.52$, and (c) $\sqrt{3}E_{x}\delta_{HL}/12 (\kappa B_0)^2 = 25.2$, with $\kappa B_0 = (\delta_{HL}^2v_{\rm F}/96 W)^{1/3}$.
    (d--f) Fraunhofer-like patterns of critical current as a function of a small additional out-of-plane magnetic field $B_z$, giving the total flux $\Phi$ through the junction, using $\sqrt{3}E_{x}\delta_{HL}/12 (\kappa B_0)^2= 2.52$ [see (b)].
    The directions of the in-plane fields are indicated in (b):
    (d) along $\hat{y}$, (e) along $\hat{x}$, and (f) diagonally along $(\hat{x} +\hat{y})/\sqrt 2$. For all plots the aspect ratio is set to $L/W = 10$, the scale of the critical current is described by $I_{\text{max}} = I_0 \pi L/W$, and $\Phi_0 = h/2e$ is the flux quantum.}
    \label{fig:smallRashba}
\end{figure}

The final limit we can consider is that of vanishing spin--orbit coupling, $\alpha_{\rm R} \to 0$.
In this case the surviving coupling terms are $H^{(2)}_{1,2}$ and $ H^{(3)}_{3,0}$, yielding the total field
\begin{align}
    \beta_+(\theta) = \frac{-4 E_{\rm Z}}{\delta_{HL}}\left( \sqrt{3}E_{x} e^{i(2\theta+\phi_B)} + \frac{12 E_{\rm Z}^2}{\delta_{HL}} e^{i3\phi_B} \right).\label{eq:betanoSOI}
\end{align}
In Fig.~\ref{fig:smallRashba} we show the dependence of the resulting critical current on the applied magnetic field, for different ratios of $E_{x}\delta_{HL}/E_{\rm Z}^2$.
In the top row of panels we plot the critical current as a function of the in-plane field, where we have used (a) $\sqrt{3}E_{x}\delta_{HL}/12 (\kappa B_0)^2 = 0.252$, (b)  $\sqrt{3}E_{x}\delta_{HL}/12 (\kappa B_0)^2= 2.52$, and (c)$\sqrt{3}E_{x}\delta_{HL}/12 (\kappa B_0)^2 = 25.2$.
We defined again $\kappa B_0 = (\delta_{HL}^2v_{\rm F}/96 W)^{1/3}$ and used an aspect ratio of $L/W = 10$ for the junction.
In the bottom row we added for reference the Fraunhofer-like patterns of critical current that emerge when a small perpendicular magnetic field $B_z$ is added in the intermediate case of (b).
The direction of each ${\bf B}_\parallel$ is indicated with a dashed line in (b): the in-plane magnetic field in (d) points along $\hat{y}$ (red dashed line), (e) along $\hat{x}$ (green dashed line), and (f) along $ (\hat{x} +\hat{y})/\sqrt 2$ (blue dashed line).

In the two limiting cases, for small and large $E_{x}\delta_{HL} / (\kappa B_0)^2$ [Fig.~\ref{fig:smallRashba}(a) and (c), respectively], we see, as expected, patterns that are similar to those for the Zeeman-dominated cases investigated above, showing the same circularly symmetric patterns with a ``linear'' dependence on $B_\parallel$ in the former case and the ``cubic'' dependence in the latter [cf.~Figs.~\ref{fig:2DEGEQ}(a) and \ref{fig:2DEGdist}(a)].
In the intermediate case (b), where $\sqrt{3}E_{x}\delta_{HL} = 12 (\kappa B_0)^2$, we expect a crossover from the pattern seen in (c) at small fields (where $E_x$ dominates) to the pattern of (a) for large fields (where $E_{\rm Z}$ dominates).
Closer inspection of Fig.~\ref{fig:smallRashba}(b) seems to confirm this behavior, also for increasing field; when we plot for even larger $B_\parallel \gtrsim 5\, B_0$ the pattern indeed becomes circularly symmetric again.

Deriving approximate expressions for the Cooper-pair propagator in the two limits again,
\begin{align}
    C({\bf r}) \approx \frac{K}{r^2} \times \begin{cases}
     \displaystyle
     \cos\left(a r \right)  & \text{for } E_{x}\delta_{HL} \gg E_{\rm Z}^2  \\
     \displaystyle
     \cos\left(b r \right)  & \text{for }  E_{x}\delta_{HL} \ll E_{\rm Z}^2
\end{cases},
\end{align}
with $a = (\sqrt 3 E_x \delta_{HL}/12 \kappa^2 B_0^2)(B_\parallel/B_0W)$ and $b  = (B_\parallel/B_0)^3(1/W)$, we can again arrive at approximate analytic expressions in the limit $L\gg W$.
In both cases one finds again the functional form of Eq.~(\ref{eq:absValIntegral}), where one now has to use ${\alpha}_{a} = (\sqrt 3 E_x \delta_{HL}/12 \kappa^2 B_0^2)(B_\parallel/B_0)$ and ${\alpha}_{b}  = (B_\parallel/B_0)^3$, respectively, as expected.

The additional structure observed in Fig.~\ref{fig:smallRashba}(b) around $B_\parallel \approx B_0$  can be understood from considering the propagator exactly at the point where $\sqrt{3}E_{x}\delta_{HL} = 12 (\kappa B_\parallel)^2$, where one finds
\begin{equation}
    C({\bf r}) = \frac{K}{r^2} \cos\left({\bf c} \cdot {\bf r} \right),
\end{equation}
with ${\bf c} = 4(3)^{1/4}\sqrt{E_x^3/\delta_{HL}v_{\rm F}^2}\,\hat{b}_\parallel$.
This leads straightforwardly to an analytic expression for the critical current in the limit $L \gg W$,
\begin{align}
    I_c \approx {} & {}
    I_0 \frac{\pi L}{W}\, e^{-|\gamma \sin \phi_B|}|\cos( \gamma \cos \phi_B)|,
\label{eq:dotProdIntegral4}
\end{align}
with $\gamma = 4(3)^{1/4}\sqrt{E_x^3W^2/\delta_{HL}v_{\rm F}^2}$, where we again emphasize that this expression is derived for the specific field strength where $\sqrt{3}E_{x}\delta_{HL} = 12 (\kappa B_\parallel)^2$.
This indicates that there can be an intermediate regime, where the pattern of $I_c({\bf B}_\parallel)$ does not look like a set of concentric high-current rings but only has significant supercurrent flowing when ${\bf B}_\parallel$ is oriented along $\pm \hat x$.
This is indeed consistent with the features we observe in Fig.~\ref{fig:smallRashba}(b) around $B_\parallel \approx 1.5\,B_0$.
This behavior can be qualitatively understood from considering Eq.~(\ref{eq:betanoSOI}) again:
When $\sqrt{3}E_{x}\delta_{HL} = 12 (\kappa B_\parallel)^2$ the two competing terms have exactly the same magnitude.
Most carriers that contribute to the current will propagate approximately in the $x$-direction where $\theta = 0,\pi$.
For those carriers, the total coupling thus becomes $\propto e^{i\phi_B} + e^{i3\phi_B}$, which indeed vanishes for fields along $\pm \hat y$: in those directions the coupling terms $H^{(1)}_{1,2}$ and $H^{(2)}_{3,0}$ interfere destructively.

\subsection{Discussion}

The results presented above could be useful for characterizing the effective spin physics within the heavy-hole subspace in a two-dimensional hole gas in experiment. Comparing a measured pattern of critical current as a function of in-plane field qualitatively with the patterns for the limiting and intermediate cases we present above could give an indication of the dominating spin-mixing mechanism in the heavy-hole subspace.

First and foremost, the largest amount of information can be gained if one observes a transition from one pattern to another upon increasing the magnitude of the field, e.g., from ``lobes'' to rings, as this can yield quantitative information about the ratio of the relevant terms in the Hamiltonian.
Most of the intermediate patterns we show above, Fig.~\ref{fig:2DEGEQ}(b) ($E_{\rm Z} \sim E_{so}$ for large $\sqrt{\delta_{HL}E_x}$), Fig.~\ref{fig:2DEGdist}(b--d) ($E_{\rm Z} \sim E_{so}$ for small $\sqrt{\delta_{HL}E_x}$), and Fig.~\ref{fig:smallRashba}(b) ($E_{\rm Z} \sim \sqrt{\delta_{HL}E_x}$ for $E_{so}=0$), look qualitatively different in terms of the direction along which the lobes appear, the periodicity of the lobes, and the order of the transitions from lobes to rings.
A critical-current pattern measured in a system that happens to be in one of the intermediate regimes can thus often be unambiguously connected to a parameter regime in our theory.
For instance, lobes appearing along the $x$-direction [Fig.~\ref{fig:smallRashba}(b)] are always an indication of negligible spin--orbit coupling and the intermediate regime $E_{\rm Z} \sim \sqrt{\delta_{HL}E_x}$.

In general, the more information about the system is available, the more precise conclusions one can potentially draw from a comparison to our theory.
For instance, if it is known that the spin--orbit coupling is non negligible and that the Zeeman coupling dominates over the SOI, then the periodicity of the oscillations as a function of in-plane field can reveal information about the HH--LH splitting:
Linearly spaced isotropic oscillations [Fig.~\ref{fig:2DEGEQ}(a)] show that one is in the large HH--LH splitting limit, where Eq.~\eqref{eq:ineqlargedelta} holds, while ``cubicly'' spaced isotropic oscillations [Fig.~\ref{fig:2DEGdist}(a)] suggest relatively small HH--LH splitting, where Eq.~\eqref{eq:ineqsmalldelta} holds.
The same holds if the system has negligible spin--orbit; a linear pattern [Fig.~\ref{fig:smallRashba}(c)] signals large HH--LH splitting and a cubic pattern [Fig.~\ref{fig:smallRashba}(a)] the opposite.
Equally spaced lobes for magnetic fields along the junction [Figs.~\ref{fig:2DEGEQ}(c) and \ref{fig:2DEGdist}(e)] are always a sign of strong Rashba-type spin--orbit coupling, similar to the electronic case~\cite{Hart2017}.

Comparing the regimes we consider above with realistic experimental parameters, we see that often one will be in a situation where $\sqrt{E_x \delta_{HL}} \gg E_{\rm Z},E_{so}$, i.e., the one considered in Secs.~\ref{sec:largeHH-LH} and \ref{sec:noSOI}:
Focusing for example on Ge-based 2DHGs, one typically has a HH--LH splitting of $\delta_{HL} \sim 20$--$80$~meV and an ``off-diagonal Fermi energy'' of $E_x \sim 1$--10~meV, whereas the Zeeman and spin--orbit energies $E_{\rm Z}$ and $E_{so}$ are significantly smaller.
We thus believe that currently the results presented in Secs.~\ref{sec:largeHH-LH} and \ref{sec:noSOI} are the most relevant ones for experiment.
However, since all energy scales depend in a different way on $\gamma_{1,2,3}$ and the ``bare'' $g$-factor in the valence band (all material parameters) as well as the thickness and asymmetry of the quantum well (device parameters), the opposite limit considered in Sec.~\ref{sec:smallHHLH} can become relevant for other materials and/or less conventionally designed quantum wells.


\section{Conclusion}
\label{sec:conclusion}

In this paper we studied an SNS junction where the normal part consists of a two-dimensional hole gas in which only the lowest (heavy-hole) subband is populated.
We investigated the dependence of the critical current through the junction on the direction and magnitude of an applied in-plane magnetic field.
Due to the underlying $p$-type nature of the valence band, the manifestation of the in-plane Zeeman effect as well as the spin--orbit coupling inside the heavy-hole subband has an intricate structure, yielding many qualitatively different spin-mixing mechanisms that could be at play.
We present a systematic analysis of the different regimes that potentially could be reached by varying the $g$-factor, the strength of a Rashba-type spin--orbit coupling, the out-of-plane confinement length, and the heavy-hole carrier density.
Applying a semi-classical approximation for the normal region (assuming the Fermi wave length to be the smallest relevant length scale) we present a straightforward numerical method for calculating the critical current in the junction.
The simplicity of the resulting expressions allows us to derive (approximate) analytic expressions for the critical current in all limiting cases, which show good agreement with the numerical results. These results could therefore potentially serve as a tool for investigating the detailed effective spin physics within the heavy-hole subspace of a two-dimensional hole gas.

\section*{Acknowledgments}

We gratefully acknowledge financial support via NTNU's Onsager Fellowship Program.

\putbib[bib]
\end{bibunit}

\widetext
\clearpage
\begin{bibunit}
\begin{center}
    \textbf{\large Supplemental material:\\ Effects of spin--orbit coupling and in-plane Zeeman fields on the critical current\\ in two-dimensional hole gas SNS junctions}
\end{center}

\setcounter{equation}{0}
\setcounter{section}{0}
\setcounter{figure}{0}
\setcounter{table}{0}
\setcounter{page}{1}
\makeatletter
\renewcommand{\theequation}{S\arabic{equation}}
\renewcommand{\thefigure}{S\arabic{figure}}
\renewcommand{\bibnumfmt}[1]{[S#1]}
\renewcommand{\citenumfont}[1]{S#1}

\section{Derivation of Eq.~(4) in the main text}

The current in the ground state is given by
\begin{equation}
I(\phi)=\frac{2e}{\hbar}\frac{\partial F}{\partial\phi},
\end{equation}
where $\phi$ is the phase difference between the two superconductors and $F$ is the free energy, $F=-T \ln {\rm Tr}\{ e^{-H/T} \}$, where $T$ is the temperature (using $k_{\rm B} = \hbar = 1$).

Working in an interaction picture, we split the full Hamiltonian, $H = H_0 + H_t$, into the tunnel coupling term $H_t$ [see Eq.~(2) of the main text], which we will treat perturbatively, and an unperturbed part $H_0$.
In this picture all operators gain time dependence governed by $H_0$ only, and we can define a so-called S-matrix as
\begin{equation}
    {\cal S}={T}_{\tau}\exp\left\{ -\int_{0}^{\beta}d\tau'\,{H}_{t}(\tau')\right\},
\end{equation}
where $T_\tau$ is the imaginary-time time-ordering operator and $\beta = 1/T$.
From this definition it follows that
\begin{equation}
    F=F_{0}-T\ln\langle{\cal S}\rangle_{0},
\end{equation}
where $\langle\dots\rangle_{0}$ is the Gibbs statistical average over the unperturbed ground state (below we will drop the subscript 0, which is implied from now on).

It is straightforward to show (see, e.g., Chapter 15 in Ref.~\cite{Abrikosov}) that
\begin{equation}
    F = F_{0}-T(\langle{\cal S}\rangle_{{\rm con}}-1),
\end{equation}
where
\begin{equation}
    \langle{\cal S}\rangle_{{\rm con}}=1+\Xi_{1}+\Xi_{2}+\dots,
\end{equation}
is the sum over all fully connected diagrams contributing to $\langle {\cal S}\rangle$,
\begin{equation}
    \Xi_n = \frac{(-1)^n}{n!} \int_0^\beta d\tau_1 \cdots d\tau_n \langle T_\tau H_t(\tau_1)\cdots H_t(\tau_n)\rangle_{\rm con}.
\end{equation}

We are interested in the lowest-order correction that depends on the phase difference of the two superconductors, which is fourth order in the coupling Hamiltonian $H_t$,
\begin{equation}
    \Xi_{4}=\frac{1}{4!}\int_{0}^{\beta}d\tau_{1}\cdots d\tau_{4}\,\langle\hat{T}_{\tau}\hat{H}_{t}(\tau_{1})\hat{H}_{t}(\tau_{2})\hat{H}_{t}(\tau_{3})\hat{H}_{t}(\tau_{4})\rangle.
\end{equation}
We thus insert the coupling Hamiltonian, written as
\begin{equation}
    \hat{H}_{t}=\sum_{\sigma}\int dy\left[\sqrt{\frac{\lambda_l}{\pi \nu_{\rm eff}}}e^{2\pi i \varphi_l(y)} \hat{\psi}_{\sigma}^{\dagger}(0,y)\hat{\Psi}_{\sigma,L}(0,y)+\sqrt{\frac{\lambda_r}{\pi \nu_{\rm eff}}}e^{2\pi i \varphi_r(y)}\hat{\psi}_{\sigma}^{\dagger}(W,y)\hat{\Psi}_{\sigma,R}(W,y)+{\rm H.c.}\right],
\end{equation}
in terms of the coupling parameters $\lambda_{l,r}$ introduced in the main text.
Here, $\nu_{\rm eff}$ is the effective one-dimensional tunneling density of states of the superconducting contacts (assumed to be equal in the two superconductors, for simplicity), $\hat \psi_\sigma^\dagger({\bf r})$ and $\hat \Psi^\dagger_{\sigma,L(R)}({\bf r})$ are the creation operators for an electron with spin $\sigma$ at position ${\bf r}$ in the normal region and the left(right) superconducting contact, respectively, and the phases $\varphi_{l,r}(y)$ are defined through
\begin{equation}
    \varphi_l(y) - \varphi_r(y') = \phi + \frac{\pi(y+y')B_zW}{\Phi_0},
\end{equation}
with $\Phi_0 = h/2e$ the flux quantum.
These phases thus incorporate the phase difference of the two superconductors as well as the orbital effect of an out-of-plane magnetic field.
We note that with this gauge choice the field operators $\hat \Psi^{(\dagger)}({\bf r})$ for the electrons in the superconductors are no longer dependent on the superconducting phases.

We collect the contributions to $\Xi_4$ that depend on $\phi$ and apply Wick's theorem to separate them into single-particle Green functions, yielding
\begin{align}
    \Xi_{4} ={} & {}  T^{4}\frac{\lambda_l\lambda_r}{2\pi^2 \nu_{\rm eff}^2}\sum_{\substack{\sigma_{1..4} \\ k_{1..4}}}\int{dy_{1..4}}\int_{0}^{\beta}d\tau_{1..4}
    \nonumber\\
    {} & {} \times \bigg[-e^{\frac{i}{2}\Delta\varphi(y_1,y_2,y_3,y_4)}
    e^{-i\omega_{k_{1}}(\tau_{1}-\tau_{2})}{\cal G}_{eh}^{{\rm sc}}(0,y_{1},\sigma_{1};0,y_{2},\sigma_{2};i\omega_{k_{1}})
    e^{-i\omega_{k_{2}}(\tau_{3}-\tau_{4})}{\cal G}_{he}^{{\rm sc}}(W,y_{3},\sigma_{3};W,y_{4},\sigma_{4};i\omega_{k_{2}}) \nonumber\\
    {} & {} \hspace{3em} \times e^{-i\omega_{k_{3}}(\tau_{3}-\tau_{1})}{\cal G}(W,y_{3},\sigma_{3};0,y_{1},\sigma_{1};i\omega_{k_{3}}) e^{-i\omega_{k_{4}}(\tau_{4}-\tau_{2})}{\cal G}(W,y_{4},\sigma_{4};0,y_{2},\sigma_{2};i\omega_{k_{4}}) \nonumber \\
    {} & {} \hspace{1.8em}- e^{-\frac{i}{2}\Delta\varphi(y_1,y_2,y_3,y_4)}e^{-i\omega_{k_{1}}(\tau_{1}-\tau_{2})}{\cal G}_{eh}^{{\rm sc}}(W,y_{1},\sigma_{1};W,y_{2},\sigma_{2};i\omega_{k_{1}}) e^{-i\omega_{k_{2}}(\tau_{3}-\tau_{4})}{\cal G}_{he}^{{\rm sc}}(0,y_{3},\sigma_{3};0,y_{4},\sigma_{4};i\omega_{k_{2}}) \nonumber\\
    {} & {} \hspace{3em} \times e^{-i\omega_{k_{3}}(\tau_{3}-\tau_{1})}{\cal G}(0,y_{3},\sigma_{3};W,y_{1},\sigma_{1};i\omega_{k_{3}}) e^{-i\omega_{k_{4}}(\tau_{4}-\tau_{2})}{\cal G}(0,y_{4},\sigma_{4};W,y_{2},\sigma_{2};i\omega_{k_{4}})\bigg],\label{eq:xi4}
\end{align}
where $\Delta\varphi(y_1,y_2,y_3,y_4) = \frac{1}{2}\left[\varphi_l(y_1)+ \varphi_l(y_2)-\varphi_r(y_3)-\varphi_r(y_4)\right]$, there are sums over fermionic Matsubara frequencies $\omega_k = (2k+1)\pi T$, and we used the standard definitions
\begin{align}
    -\langle {T}_{\tau}\hat{\Psi}_{\sigma_{1}}(x_{1},y_{1};\tau_{1})\hat{\Psi}_{\sigma_{2}}(x_{2},y_{2};\tau_{2})\rangle = {} & {} T\sum_{k}e^{-i\omega_{k}(\tau_{1}-\tau_{2})}{\cal G}_{eh}^{{\rm sc}}(x_{1},y_{1},\sigma_{1};x_{2},y_{2},\sigma_{2};i\omega_{k}),\\
    -\langle {T}_{\tau}\hat{\Psi}^\dagger_{\sigma_{1}}(x_{1},y_{1};\tau_{1})\hat{\Psi}^\dagger_{\sigma_{2}}(x_{2},y_{2};\tau_{2})\rangle = {} & {} T\sum_{k}e^{-i\omega_{k}(\tau_{1}-\tau_{2})}{\cal G}_{he}^{{\rm sc}}(x_{1},y_{1},\sigma_{1};x_{2},y_{2},\sigma_{2};i\omega_{k}),\\
    -\langle {T}_{\tau}\hat{\psi}_{\sigma_{1}}(x_{1},y_{1};\tau_{1})\hat{\psi}^\dagger_{\sigma_{2}}(x_{2},y_{2};\tau_{2})\rangle = {} & {} T\sum_{k}e^{-i\omega_{k}(\tau_{1}-\tau_{2})}{\cal G}(x_{1},y_{1},\sigma_{1};x_{2},y_{2},\sigma_{2};i\omega_{k}),
\end{align}
where we dropped the subscripts $L,R$~\cite{fn1}.
We then assume that the Andreev reflection processes described by ${\cal G}^{\rm sc}_{eh}$ and ${\cal G}^{\rm sc}_{he}$ in Eq.~(\ref{eq:xi4}) are local and energy-independent, which we do via the substitutions
\begin{eqnarray}
{\cal G}_{eh}^{{\rm sc}}(0,y_{1},\sigma_{1};0,y_{2},\sigma_{2};i\omega_{k}) & = &
\pi\nu_{\rm eff}\sigma_{1}\delta(y_{1}-y_{2}) \delta_{\sigma_{2},\bar{\sigma}_{1}},\\
{\cal G}_{eh}^{{\rm sc}}(W,y_{1},\sigma_{1};W,y_{2},\sigma_{2};i\omega_{k}) & = & 	\pi\nu_{\rm eff}\sigma_{1}\delta(y_{1}-y_{2}) \delta_{\sigma_{2},\bar{\sigma}_{1}},\\
{\cal G}_{he}^{{\rm sc}}(0,y_{1},\sigma_{1};0,y_{2},\sigma_{2};i\omega_{k}) & = & -	\pi\nu_{\rm eff}\sigma_{1}\delta(y_{1}-y_{2}) \delta_{\sigma_{2},\bar{\sigma}_{1}},\\
{\cal G}_{he}^{{\rm sc}}(W,y_{1},\sigma_{1};W,y_{2},\sigma_{2};i\omega_{k}) & = & -	\pi\nu_{\rm eff}\sigma_{1}\delta(y_{1}-y_{2}) \delta_{\sigma_{2},\bar{\sigma}_{1}}.
\end{eqnarray}

After some rearrangements and using the relation ${\cal G}(W,y',\sigma';0,y,\sigma;i\omega_{k}) = {\cal G}(0,y,\sigma;W,y',\sigma';-i\omega_{k})^*$ this yields the expression
\begin{align}
\Xi_4 = {} & {} \lambda_l\lambda_r \int dy\,dy'\, {\rm Re} \bigg\{
e^{i[\varphi_l(y) - \varphi_r(y')]}
\sum_{k} {\rm Tr} \big[
\bar{\cal G}(W,y';0,y;i\omega_{k}) \sigma_y \bar{\cal G}(W,y';0,y;-i\omega_{k})^T\sigma_y \big] \bigg\},
\end{align}
where we introduced the matrix notation
\begin{equation}
\bar{\cal G} (x',y';x,y;i\omega_k) = \begin{pmatrix}
{\cal G}_{\uparrow\uparrow}(x',y';x,y;i\omega_k) & {\cal G}_{\uparrow\downarrow}(x',y';x,y;i\omega_k)  \\
{\cal G}_{\downarrow\uparrow}(x',y';x,y;i\omega_k) & {\cal G}_{\downarrow\downarrow}(x',y';x,y;i\omega_k)
\end{pmatrix}.
\end{equation}
Using the fact that $F^{(4)} = -T \Xi_4$, we arrive at Eq.~(4) of the main text.

\section{Derivation of Eq.~(22) in the main text}

In this Section we will show how we derive Eq.~(22) from Eq.~(4) in the main text; the derivation follows the same approach as the one outlined in Ref.~\cite{Hart2017}.
We start by assuming translational invariance within the normal region of the junction, which means that we can write the correlation function (6) in the main text as
\begin{equation}
        C({\bf r})= \frac{T}{2} \sum_{k} {\rm Tr} \big[
        \bar{\cal G}({\bf r};i\omega_{k})\sigma_y
        \bar{\cal G}({\bf r};-i\omega_{k})^T\sigma_y \big],
\end{equation}
where ${\bf r}$ is now the distance vector ${\bf r}' - {\bf r}$ in terms of the coordinates used above.
We then convert the sum over Matsubara frequencies to an integral in the complex plane, which yields (assuming zero temperature for simplicity)
\begin{align}
    C({\bf r}) = {} & {} \frac{1}{4\pi i} \int_0^\infty d\omega\, 
    {\rm Tr} \Big[ \bar G^R({\bf r};\omega) \sigma_y \bar G^A({\bf r};-\omega)^T\sigma_y -  \bar G^A({\bf r};\omega) \sigma_y \bar G^R({\bf r};-\omega)^T\sigma_y\Big],
\end{align}
in terms of the retarded and advanced Green function matrices $\bar G^{A,R}$.
We rewrite these matrices using the two ${\bf k}$-dependent spin eigenstates $\ket{\lambda_{\bf k}}$,
\begin{align}
    \bar G^{R,A}({\bf r};\omega) = \frac{1}{(2\pi)^2} \int d{\bf k}\, e^{i{\bf k}\cdot{\bf r}}\sum_{\lambda_{{\bf k}}} \frac{\ket{\lambda_{{\bf k}}}\bra{\lambda_{{\bf k}}}}{\omega - \epsilon_{{\bf k}\lambda} \pm i\eta},
\end{align}
with $\eta = 0^+$ and $\epsilon_{{\bf k}\lambda}$ the eigenenergy of the state $\ket{\lambda_{\bf k}}$.
After some manipulation this yields
\begin{align}
    C({\bf r}) = \frac{1}{2(2\pi)^4} 
    \int d{\bf k}\,d{\bf k}'\,\sum_{\lambda_{{\bf k}},\lambda_{{\bf k}'}}   e^{i({\bf k}+{\bf k}')\cdot{\bf r}} 
     \bigg(\frac{\theta(-\epsilon_{{\bf k}'\lambda})}{-\epsilon_{{\bf k}'\lambda}-\epsilon_{{\bf k}\lambda}} + \frac{\theta( \epsilon_{{\bf k}\lambda})}{ \epsilon_{{\bf k}\lambda}+\epsilon_{{\bf k}'\lambda}} \bigg) \text{Tr} \Big[ \ket{\lambda_{{\bf k}}}\bra{\lambda_{{\bf k}}} \sigma_y \ket{\lambda_{{\bf k}'}}\bra{\lambda_{{\bf k}'}}^T\sigma_y\Big],
\end{align}
where $\theta(x)$ is the Heaviside step function.
We rewrite this expression in terms of an integral over two variables $\epsilon$ and $\epsilon'$, the limits of which incorporate the effect of the step functions,
\begin{align}
    C({\bf r}) = {} & {} \frac{1}{2(2\pi)^4} 
    \int d{\bf k}\,d{\bf k}' \sum_{\lambda_{{\bf k}},\lambda_{{\bf k}'}}   e^{i({\bf k}+{\bf k}')\cdot{\bf r}} \,
     \text{Tr} \Big[ \ket{\lambda_{{\bf k}}}\bra{\lambda_{{\bf k}}} \sigma_y \ket{\lambda_{{\bf k}'}}\bra{\lambda_{{\bf k}'}}^T\sigma_y \Big] \nonumber\\
    {} & {} \hspace{1em}\times \Bigg(
    \int_0^\infty d\epsilon \int d\epsilon'
    \frac{\delta(\epsilon'-\epsilon_{{\bf k}'\lambda})\delta(\epsilon - \epsilon_{{\bf k}\lambda})} {\epsilon'+\epsilon} -
    \int d\epsilon \int_{-\infty}^0 d\epsilon' \frac{\delta(\epsilon'-\epsilon_{{\bf k}'\lambda})\delta(\epsilon - \epsilon_{{\bf k}\lambda})} {\epsilon'+\epsilon} 
    \Bigg).
\end{align}
We see that when $\epsilon >0$ and $\epsilon' < 0$ the two terms on the second line cancel, so
\begin{align}
    \hspace{-1em} C({\bf r}) = {} & {} \frac{1}{2(2\pi)^4} 
    \int_0^{\infty} d\epsilon \int_0^{\infty} d\epsilon' 
    \int d{\bf k}\,d{\bf k}' \sum_{\lambda_{{\bf k}},\lambda_{{\bf k}'}}   e^{i({\bf k}+{\bf k}')\cdot{\bf r}} \,
    \text{Tr} \Big[ \ket{\lambda_{{\bf k}}}\bra{\lambda_{{\bf k}}} \sigma_y \ket{\lambda_{{\bf k}'}}\bra{\lambda_{{\bf k}'}}^T\sigma_y \Big] \nonumber\\
    {} & {} \hspace{1em}  \times \Bigg(
    \frac{\delta(\epsilon'-\epsilon_{{\bf k}'\lambda})\delta(\epsilon - \epsilon_{{\bf k}\lambda})} {\epsilon'+\epsilon} +
    \frac{\delta(-\epsilon'-\epsilon_{{\bf k}'\lambda})\delta(-\epsilon - \epsilon_{{\bf k}\lambda})} {\epsilon'+\epsilon} 
    \Bigg).
\end{align}
We now define a propagator
\begin{align}\label{eq:spectral}
    \bar g({\bf r},\epsilon) = \frac{1}{(2\pi)^2} \int d{\bf k} \, \sum_{\lambda_{{\bf k}}} e^{i{\bf k}\cdot{\bf r}} \delta(\epsilon - \epsilon_{{\bf k}\lambda})\ket{\lambda_{\bf k}}\bra{\lambda_{\bf k}},
\end{align}
which finally yields the expression 
\begin{align}
    C({\bf r}) = {} & {} \int_0^{\infty} d\epsilon \int_0^{\infty} d\epsilon' \frac{{\rm Tr} \big[\bar g({\bf r},\epsilon) \sigma_y \bar g({\bf r},\epsilon')^T \sigma_y + \bar g({\bf r},-\epsilon) \sigma_y \bar g({\bf r},-\epsilon')^T \sigma_y \big]}{2(\epsilon + \epsilon')},
\end{align}
which is Eq.~(22) of the main text.

\section{\label{semiclassical}Semi-classical approximation}

We now specify the Hamiltonian for the normal region [see Eq.~(20) of the main text],
\begin{equation}
    H_{{\bf k}} = v_{\rm F}(k-k_{\rm F}) + {\boldsymbol \beta}(\theta)\cdot {\boldsymbol \sigma},
\end{equation}
where $v_{\rm F}$ is the Fermi velocity, $k_{\rm F}$ is the Fermi momentum, ${\boldsymbol\sigma}$ is the vector of the three Pauli spin matrices, $\theta$ is the in-plane angle of the wave vector ${\bf k}$, and $\boldsymbol\beta$ is the momentum-dependent effective field acting on the spin of the propagating carriers.
Introducing the projection operator $P^{\lambda_{\bf k}} = \ket{\lambda_{\bf k}}\bra{\lambda_{\bf k}} = \frac{1}{2} [\mathbb{1} + \lambda \hat{\boldsymbol \beta}(\theta)\cdot {\boldsymbol \sigma}]$, where $\hat{\boldsymbol \beta}(\theta) = {\boldsymbol \beta}(\theta)/|{\boldsymbol \beta}(\theta)|$, and using that $\epsilon_{{\bf k}\lambda} = v_{\rm F}(k-k_{\rm F}) + \lambda |\boldsymbol \beta(\theta)|$, we can write
\begin{equation}
        \bar g({\bf r},\epsilon) \approx 
        \frac{k_{\rm F}}{(2\pi)^{2}v_{\rm F}}
         \int_0^{2\pi} d\theta\sum_{\lambda_{\bf k}} 
         e^{i\left( [ \epsilon - \lambda|\boldsymbol\beta(\theta)|]/v_{\rm F} + k_{\rm F}\right)r\cos \theta} P^{\lambda_{\bf k}},\label{eq:gr}
\end{equation}
where $r = |{\bf r}|$ and we have assumed that $(\epsilon - \lambda|\boldsymbol\beta(\theta)|)/v_{\rm F} \ll k_{\rm F}$, which amounts to implying that all the relevant dynamics happen close to the Fermi energy.
The in-plane angle $\theta$ in (\ref{eq:gr}) is now defined to be $\theta = 0$ in the direction of ${\bf r}$.
Using these approximations, ${\bf k}$ in $\lambda_{\bf k}$ is approximated as $|{\bf k}| \approx k_{\rm F}$, so $\lambda_{\bf k}$ is only a function of the direction of ${\bf k}$.

We thus denote the projector from now on as $P^{\lambda_{\theta}}$ and write
\begin{equation}
        \bar g({\bf r},\epsilon) =
        \frac{k_{\rm F}}{(2\pi)^{2}v_{\rm F}}
         \int_0^{2\pi} d\theta \,\sum_{\lambda_{\bf k}} 
        e^{i f(\theta) r\cos \theta} P^{\lambda_{\theta}},
\end{equation}
with the shorthand notation $f(\theta) = (\epsilon - \lambda|\boldsymbol\beta(\theta)|)/v_{\rm F} + k_{\rm F}$.
We now use that
\begin{equation}
    e^{iz \cos(\theta)}=\sum_n i^n J_n(z)e^{in\theta}, 
\end{equation}
where $J_n(z)$ are Bessel functions of the first kind.
This allows us to write
\begin{equation}
        \bar g({\bf r},\epsilon) =
        \frac{k_{\rm F}}{(2\pi)^{2}v_{\rm F}}
        \int_0^{2\pi} d\theta \,\sum_{\lambda_{\bf k}} \sum_{n} 
        i^n J_n[f(\theta)r]e^{in\theta} P^{\lambda_{\theta}}.
\end{equation}
Using again that $(\epsilon - \lambda|\boldsymbol\beta(\theta)|)/v_{\rm F} \ll k_{\rm F}$ and the semi-classical limit $k_{\rm F}r \gg 1$, we use the asymptotic limit of the Bessel functions
$J_n(z) {\approx} \sqrt{2/\pi z}\cos(-z+\frac{1}{4}\pi+\frac{1}{2}n\pi)$ for large $z$,
\begin{align}
        \bar g({\bf r},\epsilon) {} & {} \approx
        \frac{k_{\rm F}}{(2\pi)^{2}v_{\rm F}}
        \int_0^{2\pi} d\theta \,\sum_{\lambda_{\bf k}} \sum_{n} 
        \sqrt{\frac{2}{\pi k_{\rm F}r}} \cos \left( -f(\theta)r +\frac{\pi}{4} + \frac{n\pi}{2} \right)
        e^{in(\theta+\pi/2)} P^{\lambda_{\theta}} \nonumber\\
        {} & {} = \frac{k_{\rm F}}{(2\pi)^{2}v_{\rm F}}
        \int_0^{2\pi} d\theta \,\sum_{\lambda_{\bf k}} \sum_{l}
        \sqrt{\frac{1}{2\pi k_{\rm F}r}}
        \left[e^{i\left(f(\theta)r-\frac{\pi}{4}\right)}e^{in\theta} + e^{i\left(-f(\theta)r+\frac{\pi}{4}\right)}e^{in(\theta+\pi)}\right]
        P^{\lambda_{\theta}}.
\end{align}
We then make use of the fact that $\sum_n e^{inx} = 2\pi \delta(x)$, yielding
\begin{align}
        \bar g({\bf r},\epsilon) {} & {} = \frac{\sqrt{k_{\rm F}}}{(2\pi)^{3/2}v_{\rm F}\sqrt r}
        \sum_{\lambda} 
        \left[e^{i\left(f(0)r-\frac{\pi}{4}\right)}P^{\lambda_{0}} + e^{i\left(-f(\pi)r+\frac{\pi}{4}\right)}P^{\lambda_{\pi}}
        \right] \nonumber\\
        {} & {} = \frac{\sqrt{k_{\rm F}}}{(2\pi)^{3/2}v_{\rm F}\sqrt r}
        \sum_{\lambda} 
        \left[e^{i[(\epsilon - \lambda|\boldsymbol\beta(0)|)/v_{\rm F} + k_{\rm F}]r-i\pi/4}P^{\lambda_{0}} +
        e^{-i[(\epsilon - \lambda|\boldsymbol\beta(\pi)|)/v_{\rm F} + k_{\rm F}]r+i\pi/4}P^{\lambda_\pi}
        \right],
\end{align}
where $\lambda_{0,\pi}$ with $\lambda = \pm$ thus label the eigenstates parallel and antiparallel to ${\bf r}$, respectively.

We now insert this semi-classical result into the expression for the Cooper-pair propagator $C({\bf r})$ and use that
\begin{equation}
    \int_{0}^\infty d\epsilon \int_{0}^\infty d\epsilon^\prime \,
    \frac{e^{\pm i(\epsilon - \epsilon^\prime)a}}{\epsilon + \epsilon^\prime} = \frac{\pi}{2a},
\end{equation}
which yields
\begin{equation}
    C({\bf r}) = 
    \frac{k_{\rm F}}{(4\pi)^2v_{\rm F}r^2} 
    \text{Tr}\left\{
    \sum_{\lambda,\lambda^\prime}\left[P^{\lambda_0}\sigma_y (P^{\lambda_\pi})^T \sigma_y+ P^{\lambda^\prime_\pi}\sigma_y (P^{\lambda^\prime_0})^T \sigma_y\right]
    \right\}
    e^{-i(\lambda|\boldsymbol\beta(0)|-\lambda^\prime|\boldsymbol\beta(\pi)|)r/v_{\rm F}}.
\end{equation}

We now use that $P^{\lambda_{\bf k}} = \frac{1}{2}[\mathbb{1} + \lambda \hat{\boldsymbol\beta}(\theta) \cdot {\boldsymbol\sigma}]$ and $\sigma_y (P^{\lambda_{\bf k}})^T \sigma_y = \frac{1}{2} [\mathbb{1} - \lambda \hat{\boldsymbol\beta}(\theta) \cdot {\boldsymbol\sigma}]$ which finally yields
\begin{equation}
    C({\bf r}) = 
    \frac{k_{\rm F}}{(4\pi)^2v_{\rm F}r^2} 
    \sum_{\lambda,\lambda^\prime}\left[ 1 - \lambda \lambda^\prime \,\hat{\boldsymbol\beta}(0) \cdot \hat{\boldsymbol\beta}(\theta) 
    \right]
    e^{-i(\lambda|\boldsymbol\beta(0)|-\lambda^\prime|\boldsymbol\beta(\pi)|)r/v_{\rm F}},
\end{equation}
which reduces to Eq.~(24) in the main text after summing over $\lambda, \lambda' = \pm$.
In the main text we denoted the directions parallel and antiparallel to ${\bf r}$ with $\theta$ and $\bar\theta$, respectively.

\section{Approximate analytic solutions of Eq.~(30)}

\subsection{$C({\bf r}) = K \cos(|{\bf d}|r)/r^2$}

We first consider the integral (30) in the limit where
\begin{equation}
    C({\bf r}) = K \cos(|{\bf d}|r)/r^2,
\end{equation}
which is first discussed in Sec.~III.A.
After switching to sum and difference coordinates, $\sigma = \frac{1}{2}(y+y')$ and $\tau = y-y'$, the integral over $\sigma$ can be easily performed.
Substituting $\rho = \sqrt{(\tau/W)^2+1}$ then yields
\begin{equation}
    I_c({\bf B}_\parallel) = 2 I_0 \int_1^q d\rho \, \frac{\cos (\alpha \rho)}{\rho}
    \left( \frac{L}{W\sqrt{\rho^2-1}} - 1 \right),\label{eq:intappr1Supp}
\end{equation}
where $I_0 = 4K|\lambda_l\lambda_r|$ and $\alpha = B_\parallel/B_0$.
A solution in the limit $q = \sqrt{1+(L/W)^2} \to \infty$ is presented in the main text, but for general $L/W$ the first term in (\ref{eq:intappr1Supp}) is not analytically solvable.
To arrive at an approximate solution we substitute
\begin{equation}
    \frac{1}{\rho\sqrt{\rho^2 - 1}} \quad \to  \quad \frac{\rho^2-\sqrt{2\rho-2}}{\rho^2\sqrt{2\rho-2}}e^{5(1-\rho)/2} + \frac{1}{\rho^2},
\end{equation}
which is accurate within a few percent for all $\rho > 1$.
This yields the approximate result
\begin{align}
    I_c({\bf B}_\parallel) \approx {} & {} I_0 \frac{2L}{W}
    \left( \alpha[ {\rm Si}(\alpha) - {\rm Si}(q\alpha) ]
    + \frac{e^{5(1-q)/2} -1}{q}\cos(q\alpha)
    + e^{5/2}{\rm Re} \left[ e^\zeta \sqrt{\frac{\pi}{-2\zeta}} {\rm erf} (\sqrt{(1-q)\zeta})
    + \zeta \big({\rm Ei}[\zeta] - {\rm Ei}[q\zeta] \big)\right]\right)\nonumber\\
    {} & {} + 2I_0 \left[ {\rm Ci}(\alpha) - {\rm Ci}(q\alpha) \right],\label{eq:intres1}
\end{align}
where ${\rm Si}(x)$ and ${\rm Ci}(x)$ are the sine and cosine integral, respectively, ${\rm Ei}(x)$ is the exponential integral function, ${\rm erf}(x)$ is the error function, and we introduced the parameter $\zeta = -\frac{5}{2} + i\alpha$ for concise notation.

\subsection{$C({\bf r}) = K \cos(\boldsymbol\kappa \cdot {\bf r})/r^2$}

The other limiting form the Cooper pair propagator takes is
\begin{equation}
    C({\bf r}) = K \frac{\cos(\boldsymbol\kappa \cdot {\bf r})}{r^2},
\end{equation}
where $\boldsymbol\kappa$ can be $\boldsymbol\kappa \propto \hat z \times {\bf B}_\parallel$ (as in Sec.~III.A and III.B) or $\boldsymbol\kappa \propto {\bf B}_\parallel$ (as in Sec.~III.C).
In the limit of large $L/W$ one can approximate
\begin{equation}
    I_c({\bf B}_\parallel) \approx
    I_0 \int_0^L dy' \int_{-\infty}^{\infty} dy\, \frac{\cos[\kappa_x L + \kappa_y(y-y')]}{L^2 + (y-y')^2} = I_0 \frac{\pi L}{W} e^{-|\kappa_y|W}\cos(\kappa_x W).
\end{equation}
For general $L/W$ the original integral (i.e., without setting the limits of the second integral to $\pm \infty$) can be (quasi)analytically solved, resulting in
\begin{align}
    I_c({\bf B}_\parallel) = I_0 \cos(\kappa_x W) {\rm Re} \Big[ {} & {} 
    e^{|\kappa_y W|} \left( \pi \omega + {\rm Ei}\big[-|\kappa_y W|\big] - (1+i\omega ) {\rm Ei}\big[-(1+i\omega)|\kappa_y W|\big]\right) \nonumber\\
    {} & {} \hspace{1em} + e^{-|\kappa_y W|} \left( \pi \omega + {\rm Ei}\big[|\kappa_y W|\big] - (1+i\omega ) {\rm Ei}\big[(1+i\omega)|\kappa_y W|\big]\right) \Big],\label{eq:intres2}
\end{align}
where $\omega = L/W$ is the aspect ratio of the junction and ${\rm Ei}(x)$ is again the exponential integral function.

\end{bibunit}

\begin{thebibliography}{1}
\bibitem{fn1}
The prefactor of Eq.~(\ref{eq:xi4}) is $1/4$ for the linked cluster expansion we are doing here instead of 1, when all imaginary times $\tau_n$ can be permuted (see, e.g., Chapter 3 in Ref.~\cite{mahan:book}). An additional factor 2 arises from the fact that the two superconductors are not equivalent, which yields two choices for which superconductor interacts with the normal region at $\tau_1$.
\end{thebibliography}
\end{document}